\documentclass[aps,prd,twocolumn,showpacs,showkeys,superscriptaddress]{revtex4}
\usepackage{epsfig}
\usepackage{graphicx}
\usepackage{amsmath,amssymb}

\newcommand{\bastar}{\begin{eqnarray*}}
\newcommand{\eastar}{\end{eqnarray*}}
\newskip\humongous \humongous=0pt plus 1000pt minus 1000pt

\newif\ifdtup

\relax
\newcommand{\bea}{\begin{eqnarray}}
\newcommand{\eea}{\end{eqnarray}}
\newcommand{\X}{{\vec X}}
\newcommand{\W}{{\vec W}}
\newcommand{\pro}{\partial}
\newcommand{\pd}{\partial}
\newcommand{\n}{\hat n}

\newcommand{\mn}{{\mu\nu}}
\newcommand{\oneg}{\displaystyle\frac{1}{g}}

\newcommand{\B}{{\vec B}}

\newcommand{\R}{{\vec R}}
\newcommand{\G}{{\vec G}}

\newcommand{\F}{\vec F}
\newcommand{\vH}{\vec H}
\newcommand{\vE}{\vec E}
\newcommand{\hF}{\hat F}

\newcommand{\bH}{\bar H}
\newcommand{\bE}{\bar E}

\newcommand{\A}{{\vec A}}
\newcommand{\hA}{{\hat A}}
\newcommand{\cA}{{\cal A}}
\newcommand{\cC}{{\cal C}}
\newcommand{\cD}{{\cal D}}
\newcommand{\valpha}{{\vec \alpha}}

\newcommand{\hn}{{\hat n}}
\newcommand{\hD}{{\hat D}}
\newcommand{\bD}{{\bar D}}

\newcommand{\nn}{\nonumber}

\newcommand{\Int}{\displaystyle{\int}}
\newcommand{\vc}{\vec c}

\newcommand{\bR}{\bar R}
\newcommand{\bB}{\bar B}
\newcommand{\bG}{\bar G}
\begin{document}
\title{Abelian Decomposition and Weyl Symmetric Effective 
Action of SU(3) QCD}
\bigskip
\author{Y. M. Cho}
\email{ymcho0416@gmail.com}
\affiliation{Center for Quantum Spacetime, Sogang University,
Seoul 04107, Korea}  
\affiliation{School of Physics and Astronomy,
Seoul National University, Seoul 08826, Korea}
\author{Franklin H. Cho}
\affiliation{Institute for Quantum Computing,
Department of Physics and Astronomy  \\
University of Waterloo, Ontario N2L 3G1, Canada}

\begin{abstract}
We show how to calculate the effective potential of SU(3) 
QCD which tells that the true minimum is given by the monopole 
condensation. To do this we make the gauge independent 
Weyl symmetric Abelian decomposition of the SU(3) QCD 
which decomposes the gluons to the color neutral neurons 
and the colored chromons. In the perturbative regime this decomposes the Feynman diagram in such a way that 
the conservation of color is explicit. Moreover, this shows 
the existence of two gluon jets, the neuron jet and chromon 
jet, which can be verified by experiment. In the non-perturbative regime, the decomposition puts QCD to the background field formalism and reduces the non-Abelian gauge symmetry to 
a discrete color reflection symmetry, and provides us an ideal platform to calculate the one-loop effective action of QCD. Integrating out the chromons from the Weyl symmetric Abelian decomposition of QCD gauge invariantly imposing the color 
reflection invariance, we obtain the SU(3) QCD effective 
potential which generates the stable monopole condensation 
and the mass gap. We discuss the physical implications of 
our result, in particular the possible existence of the vacuum fluctuation mode of the monopole condensation in QCD.  
\end{abstract}
\maketitle

\section{Introduction}

The color confinement problem in quantum chromodynamics 
(QCD) is one of the most challenging problems in theoretical 
physics. Two leading conjectures of the confinement mechanism
are the monopole condensation \cite{nambu,prd80,prl81} 
and the Abelian dominance \cite{thooft,prd00}. The Abelian 
dominance asserts that only the Abelian (diagonal) part of 
the gluons is responsible for the color confinement. Intuitively 
this must be true, because the non-Abelian (off-diagonal) 
part describes the colored gluons which are destined to be 
confined. Since the confined prisoner can not be the confining 
agent (the jailer), only the Abelian part can play the role of 
the confiner.

In fact we can prove this Abelian dominance rigorously. 
Theoretically we can show that the contribution of 
the non-Abelian part in the area law of the Wilson loop 
integral is negligible \cite{prd00}. Moreover, numerically  
we can confirm this in the lattice QCD \cite{kondo,cundy}.   

The problem with this conjecture is that this does not tell 
what is exactly the Abelian part of QCD, and how different 
is this from the Abelian gauge theory. More importantly, 
this does not tell how is the color confined by the Abelian 
part.     

The monopole condensation tries to explain how is the color 
confined. It has long been argued that the confinement in 
QCD can be triggered by the monopole condensation. Indeed, 
if one assumes the monopole condensation, one could 
argue that the ensuing dual Meissner effect generates 
the color confinement \cite{nambu,prd80,prl81}. Proving 
the monopole condensation, however, has been extremely 
difficult. 

A natural way to establish the monopole condensation in QCD 
is to show that the quantum fluctuation triggers a phase 
transition similar to the dimensional transmutation observed 
in massless scalar QED \cite{cole}. There have been many 
attempts to demonstrate this. Savvidy first calculated 
the effective action of SU(2) QCD integrating out the colored 
gluons in the presence of an {\it ad hoc} color magnetic 
background, and has almost ``proved'' the magnetic 
condensation known as the Savvidy vacuum \cite{savv}. 

Unfortunately, the subsequent calculation repeated by Nielsen and 
Olesen showed that the effective action has an extra imaginary 
part which destabilizes the Savvidy vacuum. This is known as 
the ``Savvidy-Nielsen-Olesen (SNO) instability'' \cite{niel,ditt,yil}. 
The origin of this instability can be traced to the tachyonic modes 
in the functional determinant of the gluon loop integral. 

But in physics we encounter tachyons when we do something 
wrong. For example, in spontaneous symmetry breaking we 
have tachyon when we choose the false vacuum. Similarly, 
in Neveu-Schwarz-Ramond (NSR) string theory we have 
the tachyonic vacuum when we do not make the theory modular 
invariant and supersymmetric with the Gliozzi-Scherk-Olive 
(GSO) projection \cite{gso,witt}. So, obviously there is something 
wrong in the instability of the SNO vacuum. The question is how 
to remove the tachyonic modes in the gluon functional determinant, 
and how to justify that. 

We emphasize, however, that the most serious defect of the SNO 
vacuum is not that it is unstable but that it is not gauge invariant. 
So even if the Savvidy vacuum were made stable, it can not be 
the QCD vacuum. Because of this Nielsen and Olesen has proposed the so-called ``Copenhagen vacuum", the randomly oriented piecewise Savvidy vacuum \cite{niel}. But one can not obtain a gauge invariant vacuum simply by randomly orienting something which is not gauge invariant.

The gauge independent Abelian decomposition plays the crucial 
role to cure this defect. It decomposes QCD gauge potential to 
the color neutral Abelian part and colored valence part gauge independently. As importantly, it tells that the Abelian potential 
is made of two parts, the non-topological (Maxwellian) Abelian 
part and the topological (Diracian) monopole par \cite{prd80,prl81}. This means that there are actually two possible magnetic backgrounds, the non-topological magnetic background 
(the Savvidy background) and topological monopole background. Moreover, we can show that only the monopole background is 
gauge invariant \cite{prd02,jhep05,prd13}. So choosing 
the monopole background in the calculation of the QCD 
effective action, we can avoid this trouble. 

The Abelian decomposition also plays the crucial role to cure 
the SNO instability. It shows that after the Abelian decomposition 
the non-Abelian gauge symmetry is reduced to a simple discrete 
symmetry called the color reflection symmetry \cite{prd80,prl81}. 
So the color reflection invariance assures the gauge invariance 
after the decomposition. This means that, integrating the colored 
gluons imposing this color reflection invariance under 
the monopole background we can calculate the QCD effective 
action gauge invariantly \cite{prd02,jhep05,prd13}. This removes 
the tachyonic modes and allows us to obtain the stable monopole condensation.

The fact that the monopole plays the crucial role in the color 
confinement is well established by now. First, using the Abelian 
decomposition we can prove that the Abelian part of the potential 
is responsible for the confining force in Wilson loop \cite{prd00}. 
This, of course, is the Abelian dominance. But we can go further, 
and establish the monopole dominance theoretically as well as 
numerically. For instance, implementing the Abelian decomposition 
on lattice we can calculate Wilson loop contribution of the full 
potential, the Abelian potential, and the monopole potential 
separately, and show that all three potentials give the same area 
law \cite{kondo,cundy}. This means that only the monopole 
potential is enough to generate the confining force.   

The lattice results, however, does not tell how the monopole 
confines the color. In the preceding papers we have shown 
how to calculate the effective action of SU(2) QCD, and demonstrated that the stable monopole condensation can 
take place which generates the mass gap and color confinement 
in SU(2) QCD \cite{prd02,jhep05,prd13}. The purpose of 
this paper is to generalize this result to the real SU(3) QCD.

The crucial step to generalize the SU(2) result to SU(3) is to 
express the Abelian decomposition in SU(3) in the Weyl 
symmetric form. One might wonder how do we have the Weyl symmetric Abelian decomposition of SU(3) QCD, when we 
have only two Abelian directions in SU(3). The trick is to 
express the two Abelian potentials to three Abelian potentials 
of the SU(2) subgroups in Weyl symmetric way. This greatly simplifies for us to calculate the SU(3) QCD effective 
action from the SU(2) QCD effective action. 

The paper is organized as follows. In section II we review 
the Abelian decomposition of SU(2) QCD for later purpose. 
In section III we discuss the Weyl symmetric Abelian 
decomposition of SU(3) QCD which greatly simplifies 
the calculation of the effective action of SU(3) QCD. In 
section IV we discuss the color reflection symmetry which 
replaces the role of the non-Abelian gauge symmetry after 
the Abelian decomposition, which plays a crucial role 
for us to implement the gauge invariance in the calculation 
of the QCD effective action. In section V we repeat 
the calculation of the one-loop effective action of SU(2) 
QCD which plays the fundamental role for the SU(3) QCD
effective action. In section VI we calculate SU(3) QCD 
effective action and demonstrate that the monopole 
condensation becomes the Weyl symmetric vacuum in SU(3) 
QCD. In particular we show that the essential features of 
SU(2) QCD, the dimensional transmutation by the monopole 
condensation which generates the mass gap remains 
the same. Finally in section VII we discuss the physical 
implications of our result. 

\section{Abelian Decomposition of SU(2) QCD: A Review} 

Before we discuss the Abelian decomposition we have to 
know why we need it. Consider the proton. The quark model 
tells that it is made of three quarks, but obviously we need 
gluon to bind them. On the other hand the quark model 
asserts that there is no ``valence"  gluon inside the proton 
which can be viewed as a constituent of proton. If so, 
what is the ``binding" gluon inside the proton, and how 
do we distinguish it from the valence gluon? 

Another motivation is the Abelian dominance. It has been
believed that the Abelian part of gluon is responsible for 
the color confinement in QCD. As we have pointed out, 
this must be true. But what is the Abelian part, and how 
do we separate it? To answer these questions we need 
the Abelian decomposition.  
   
Consider the SU(2) QCD for simplicity, and let 
$(\hn_1,\hn_2,\hn_3=\hn)$ be an arbitrary right-handed 
local orthonormal basis. To make the Abelian decomposition 
we choose any direction, for example $\hn$, to be the Abelian 
direction and impose the isometry to project out the restricted 
potential $\hA_\mu$ \cite{prd80,prl81}
\bea
&D_\mu \hn=(\pro_\mu+g\A_\mu \times) \hn=0,  \nn\\
&\A_\mu \rightarrow \hA_\mu
=A_\mu \n-\oneg \n \times \pro_\mu \n
=\cA_\mu+\cC_\mu,  \nn\\
&\cA_\mu=A_\mu \n,~~\cC_\mu=-\oneg \n \times \pro_\mu \n,
~~A_\mu=\hn \cdot \A_\mu. 
\label{ap}
\eea
This is the Abelian projection which projects out the color 
neutral (i.e., Abelian) restricted potential. 

We emphasize the followings. First, $\hA_\mu$ is precisely 
the connection which leaves the Abelian direction invariant 
under the parallel transport (which makes $\hn$ covariantly 
constant). Second, it is made of two parts, the non-topological 
(Maxwellian) $\cA_\mu$ which describes the clor neutral 
gluon (the neuron) and the topological (Diracian) $\cC_\mu$ 
which describes the non-Abelian monopole \cite{prl80}. 
Third, the decomposition is gauge independent, because 
$\n$ is arbitrary. We can rotate $\n$ to any direction and 
still get exactly the same decomposition. 

With this we have
\bea
& \hF_\mn = (F_\mn+ H_\mn)\hn, \nn\\
&F_\mn = \pro_\mu A_\nu-\pro_\nu A_\mu,  \nn\\
&H_\mn = -\oneg \n \cdot (\pro_\mu \n\times \pro_\nu \n)
=\partial_\mu C_\nu-\partial_\nu C_\mu,  \nn\\
&C_\mu=-\dfrac{1}{g} \hn_1 \cdot \pro_\mu \hn_2.
\label{rf} 
\eea 
This tells the followings. First, $\hF_\mn$ has only the Abelian 
component. Second, $\hF_\mn$ is made of two potentials, 
the non-topological $A_\mu$ and topological $C_\mu$. 
This dual structure of $\hF_\mn$ plays an important role
in the calculation of the QCD effective action, because this 
tells that there are actually two candidates of classical 
magnetic background to choose, the Savvidy background 
coming from $F_\mn$ and the monopole background 
coming from $H_\mn$ \cite{prd02,jhep05,prd13}.   

With (\ref{ap}) we can express the full SU(2) potential adding
the non-Abelian (colored) part $\X_\mu$ \cite{prd80,prl81}
\bea
&\A_\mu = \hA_\mu + \X_\mu,     \nn\\
&\X_\mu=\dfrac1g \hn \times D_\mu \hn,
~~~~\hn \cdot \X_\mu=0.
\label{adec}
\eea
Under the infinitesimal gauge transformation
$\delta \A_\mu=\oneg  D_\mu \vec \alpha$ we have 
\bea 
&\delta \hA_\mu=\dfrac1g \hD_\mu \vec \alpha, 
~~(\hD_\mu=\pro_\mu+g\hA_\mu\times),   \nn\\
&\delta \X_\mu = - \valpha \times \X_\mu.
\eea 
This tells that $\hA_\mu$ has the full SU(2) gauge degrees 
of freedom, even though it is restricted. Moreover, $\X_\mu$ 
becomes gauge covariant. This is a direct consequence of 
the fact  that the connection space (the space of all potentials) 
forms an affine space. With this $\X_\mu$ can be interpreted to 
describe the colored gluon (the chromon). This is the Abelian 
decomposition which decomposes the SU(2) gluons to one 
Abelian neuron and two colored chromons gauge independently. 
Notice that we can express the chromon by $\R_\mu$ indicating 
the color, or in the complex notation by $R_\mu$ with 
$R_\mu=(X_\mu+iX_\mu^2)/\sqrt 2$. 

This should be compared with the popular Abelian decomposition 
based on Maximal Abelian Gauge (MAG) condition \cite{thooft}.  
Here the decomposition is given by $\A_\mu=A_\mu \n
+\check X_\mu$ with $\check X_\mu=A_\mu^1 \n_1+A_\mu^2\n_2$, with the MAG condition $\check D_\mu \check X_\mu=0$. So 
the neuron and the chromon are described by the Abelian 
(diagonal) component and  the non-Abelian (off-diagonal) 
component of $\A_\mu$. 

But obviously this decomposition is not gauge independent, 
and the chromon $\check X_\mu$ does not transform 
covariantly. Worse, the topology of the non-Abelian gauge 
symmetry which plays the crucial role in QCD is completely 
neglected in this decomposition. In contrast the topological 
part plays an essential role in our Abelian decomposition. 
In fact (\ref{adec}) tells that the Abelian decomposition is 
made of three parts, the neuron, chromon, and the topological 
parts. Without the topological part we can not decompose 
the gluon to neuron and chromon gauge independently. 

With the restricted potential we can construct the restricted 
QCD (RCD) which has the full non-Abelian gauge symmetry 
but is simpler than the QCD 
\bea 
&{\cal L}_{RCD} =-\dfrac{1}{4} \hF^2_\mn
=-\dfrac{1}{4} F_\mn^2 \nn\\
&+\dfrac1{2g} F_\mn \hn \cdot (\pro_\mu \hn \times \pro_\nu \hn)
-\dfrac1{4g^2} (\pro_\mu \hn \times \pro_\nu \hn)^2,
\label{rcd}
\eea
which describes the Abelian sub-dynamics of QCD. Since 
RCD contains the non-Abelian monopole degrees explicitly, 
it provides an ideal platform for us to study the monopole
physics gauge independently.

From (\ref{adec}) we have
\bea
\vec{F}_{\mu\nu}&=&\hat F_\mn + \hD _\mu \X_\nu 
- \hD_\nu \X_\mu + g\X_\mu \times \X_\nu. 
\eea 
With this we can express QCD by
\bea 
&{\cal L}_{ECD} = -\dfrac{1}{4} \F^2_\mn
=-\dfrac{1}{4}\hF_\mn^2-\dfrac{1}{4}(\hD_\mu\X_\nu
-\hD_\nu\X_\mu)^2 \nn\\
&-\dfrac{g}{2} {\hat F}_\mn \cdot (\X_\mu \times \X_\nu)
-\dfrac{g^2}{4} (\X_\mu \times \X_\nu)^2. 
\label{2ecd} 
\eea
This is the extended QCD (ECD) which shows that QCD 
can be viewed as RCD which has the chromon as the colored 
source \cite{prd80,prl81}. 

We can easily add the quark in the Abelian decomposition,
\bea
&{\cal L}_{q} =\sum_k\bar \Psi_k (i\gamma^\mu D_\mu-m) 
\Psi_k \nn\\
&= \sum_k \Big[\bar \Psi_k (i\gamma^\mu \hD_\mu-m) \Psi_k
+\dfrac{g}{2} \X_\mu
\cdot \bar \Psi_k (\gamma^\mu \vec \tau) \Psi_k \Big], \nn\\
&\hD_\mu = \pro_\mu + \dfrac{g}{2i} {\vec \tau}\cdot \hA_\mu, 
\label{2qlag}
\eea
where $\Psi$ is the quark doublet, $m$ is the mass, and 
$k$ is the flavour index. 

Mathematically ECD is identical to QCD, but physically it 
provides a totally new meaning to QCD. It tells that QCD 
has two types of gluons, neuron and chromon, which 
plays different roles. The neuron, together with 
the topological monopole potential, provides the binding. 
On the other hand the chromon, just like the quarks, 
becomes the colored source which is destined to be 
confined \cite{prd80,prl81,prd13}.

\begin{figure}
\includegraphics[scale=0.6]{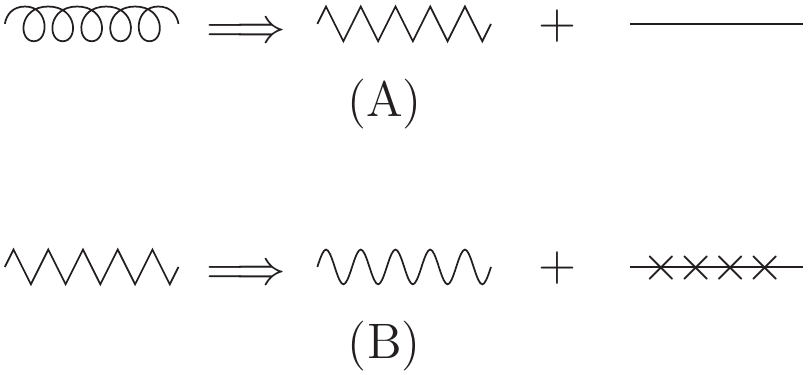}
\caption{\label{cdec} The Abelian decomposition of 
the gauge potential. In (A) it is decomposed to 
the restricted potential (kinked line) and the valence 
potential (straight line) which describes the chromon. 
In (B) the restricted potential is further decomposed 
to the Maxwell part (wiggly line) which describes 
the neuron and the Dirac part (spiked line) which 
describes the monopole.}
\end{figure}

We can express the Abelian decomposition graphically. 
This is shown in Fig. \ref{cdec}, where the gauge potential 
is decomposed to the restricted potential which has the full 
gauge degrees of freedom and the gauge covariant valence 
potential which describes the chromon in (A), and 
the restricted potential is decomposed further to 
the non-topological Maxwell part $\cA_\mu$ which 
describes the neuron and the topological Dirac part 
$\cC_\mu$ which describes the monopole in (B). 

\begin{figure}
\includegraphics[height=4cm, width=7cm]{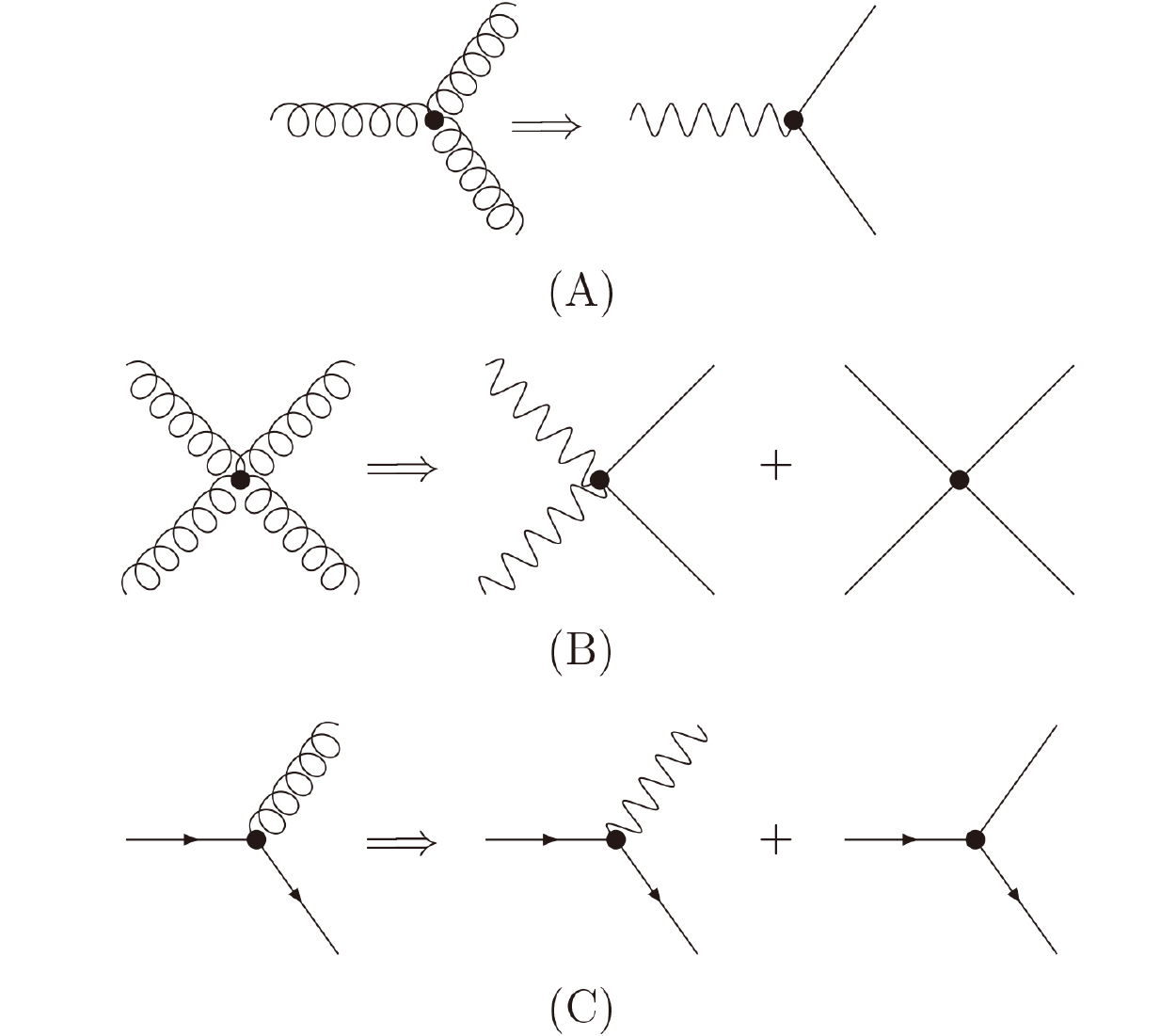}
\caption{\label{2ecdint} The decomposition of the Feynman 
diagrams in SU(2) QCD. The three-point and four-point gluon 
vertices are decomposed in (A) and (B), and the quark-gluon 
vertices are decomposed in (C). Notice that since the monopole 
is not a propagating degree it does not appear in the Feynman 
diagram.}
\end{figure}

A direct consequence of the Abelian decomposition is 
the decomposition of the Feynman diagram. Clearly (\ref{2ecd}) 
and (\ref{2qlag}) tells that the SU(2) QCD vertices can be 
decomposed to the neuron and chromon interaction. This is 
shown in Fig. \ref{2ecdint}. Notice that the conservation of 
color forbids the three-point gluon vertex made of three neurons 
or three chromons. Similarly, the four-point gluon vertex made 
of four neurons or one neuron and three chromons is forbidden. 
This is because the SU(2) QCD has only one color. Without 
the Abelian decomposition  this would have been impossible.    

At this point one might wonder why the monopole part 
does not appear in the Feynman diagram. There has been 
an assertion in the literature that the introduction of 
the Abelian direction $\n$ adds a new dynamical degree, 
and thus alters QCD \cite{fadd}. This is a gross misunderstanding 
of the Abelian decomposition. As we have emphasized, we can 
change $\n$ to any direction we like by a gauge transformation, 
for example to a trivial configuration $(0,0,1)$. This tells that 
$\n$ does not represent the dynamical degree, but the gauge 
degree. In fact, we can show explicitly that $\n$ has no equation 
of motion to satisfy, so that it can not describe a propagating
degree \cite{prd01}. This is why the monopole part does not 
appear in the Feynman diagram. 

This does not mean that $\n$ is not important. In fact it plays 
a very important role, since the non-Abelian gauge degrees of 
QCD has the non-trivial topological structure which could change 
the physics drastically. Indeed $\n$ here represents not only 
the monopole topology $\pi_2(S^2)$ but also the vacuum topology 
$\pi_3(S^3)\simeq\pi_3(S^2)$ of the SU(2) gauge theory, both of 
which are the essential characteristics of QCD \cite{prl80,plb07}. 

Moreover, it allows us to make the Abelian decomposition 
and prove that there are two types of gluons which play 
totally different roles. Most importantly it provides an ideal 
platform for us to calculate the QCD effective action 
and prove the monopole condensation. This would have 
been impossible without it. In the literature the Abelian 
decomposition has been known as the Cho decomposition, 
Cho-Duan-Ge (CDG) decomposition, or Cho-Faddeev-Niemi 
(CFN) decomposition \cite{fadd,shab,zucc}.

\section{Weyl symmetric Abelian Decomposition of SU(3) QCD}

The Abelian decomposition of SU(3) QCD is technically a bit 
more complicated but straightforward \cite{ijmpa14,prd15,prd18}. 
Since SU(3) has rank two, we have two Abelian subgroups of 
SU(3). So we need two Abelian directions to make the Abelian decomposition. 

Let $\n_i~(i=1,2,...,8)$ be the local orthonormal SU(3) basis. 
Clearly we can choose the Abelian directions to be $\n_3=\n$ 
and $\n_8=\n'$. Now make the Abelian projection by
\bea
D_\mu \n=0.
\label {3cp}
\eea
This automatically guarantees \cite{prl80}
\bea
D_\mu \n'=0,~~~\n'=\dfrac1{\sqrt 3} \n*\n.
\eea
where $*$ denotes the $d$-product. This is because SU(3) 
has two vector products, the anti-symmetric $f$-product 
and the symmetric $d$-product. This tells that we actually 
need only one $\lambda_3$-like Abelian direction to make 
the Abelian decomposition, although SU(3) has two Abelian directions.

Solving (\ref{3cp}), we have the following Abelian projection
which projects out the Abelian restricted potential,
\bea
&\A_\mu \rightarrow \hA_\mu=A_\mu \hn+A_\mu' \hn'
-\oneg \hn\times \pro_\mu \hn-\oneg \hn'\times \pro_\mu \hn' \nn\\
&=\sum_p \dfrac23 \hA_\mu^p,~~~(p=1,2,3),    \nn\\
&\hA_\mu^p=A_\mu^p \n^p-\oneg \n^p \times \pro_\mu \n^p
=\cA_\mu^p+\cC_\mu^p, \nn\\
&A_\mu^1=A_\mu,
~~~A_\mu^2=-\dfrac{1}{2}A_\mu+\dfrac{\sqrt 3}{2}A_\mu',  \nn\\
&A_\mu^3=-\dfrac{1}{2}A_\mu-\dfrac{\sqrt 3}{2}A_\mu', 
~~~\n^1=\n,   \nn\\
&\n^2=-\dfrac{1}{2} \n +\dfrac{\sqrt 3}{2} \n', 
~~~\n^3=-\dfrac{1}{2} \n -\dfrac{\sqrt 3}{2} \n',
\label{cp3}
\eea
where the sum is the sum of the three Abelian directions 
$(\n^1,\n^2,\n^3)$ of three SU(2) subgroups made of 
$(\n_1,\n_2,\n^1)$, $(\n_6,\n_7,\n^2)$, $(\n_4,\n_5,\n^3)$ \cite{r}. 
Notice the factor $2/3$ in front of $\hA_\mu^p$ in 
the $p$-summation. This is because the three SU(2) 
restricted potentials are not independent. 

What is remarkable about the above expression is that, 
although SU(3) has two Abelian directions, the restricted 
potential is expressed by the restricted potentials of three 
SU(2) subgroups in a Weyl symmetric way. The Weyl 
symmetry of SU(3) is given by the six element permutation 
group of three SU(2) subgroups, or equivalently three 
colors of SU(3). 

In general the Weyl group of SU(N) is the $N!$ elements 
permutation group of $N$ colors of SU(N), which is 
mathematically identical to the symmetric group $S_N$ 
of order $N$. And we can show that the restricted potential 
of the SU(N) QCD can be expressed by the sum of 
the restricted potentials of $N$ SU(2) subgroups in a Weyl 
symmetric way. This is very important because, as we will 
see, this allows us to calculate the SU(N) QCD effective 
action in terms of the SU(2) QCD effective action.   

Just as in the SU(2) QCD we can easily show that the SU(3) 
restricted potential has the full gauge degrees of freedom, 
and construct the SU(3) RCD made of the restricted field 
strength,
\bea
&{\cal L}_{RCD} = -\sum_p \dfrac{1}{6} (\hF_\mn^p)^2, 
\label{rcd3}
\eea
which has the full SU(3) gauge symmetry. Here again the factor 
$1/6$ comes from the fact that the three SU(2) restricted field 
strengths are not independent.   

With (\ref{cp3}) we have the Abelian decomposition of the SU(3)
gauge potential,
\bea
&\A_\mu=\hat A_\mu+\X_\mu
=\sum_p (\dfrac23 \hA_\mu^p+\W_\mu^p), \nn\\
&\X_\mu= \sum_p \W_\mu^p,  \nn\\
&\W_\mu^1= X_\mu^1 \n_1+ X_\mu^2 \n_2,
~~~\W_\mu^2=X_\mu^6 \n_6 + X_\mu^7 \n_7,  \nn\\
&\W_\mu^3= X_\mu^4 \n_4  +X_\mu^5 \n_5.
\label{cdec3}
\eea
Here again $\X_\mu$ transforms covariantly. Moreover, it 
can be decomposed to the three chromons $\W_\mu^p$ of 
the SU(2) subgroups. So we have two neurons and six (or 
three complex) chromons in SU(3) QCD. And we can identify 
$(\W_\mu^1,\W_\mu^2,\W_\mu^3)$ as the red, blue, and 
green chromons $(\R_\mu,\B_\mu,\G_\mu)$, or equivalently 
by $(R_\mu,B_\mu,G_\mu)$  with
\begin{gather} 
R_\mu=\frac{X_\mu^1+iX_\mu^2}{\sqrt 2},
~~~~B_\mu=\frac{X_\mu^6+iX_\mu^7}{\sqrt 2}, \nn\\
G_\mu=\frac{X_\mu^4+iX_\mu^5}{\sqrt 2},
\end{gather}
in the complex notation.

From (\ref{cdec3})  we have
\bea
&\hD _\mu \X_\nu=\sum_p \hD_\mu^p \W_\nu^p,
~~~\hD_\mu^p=\pro_\mu+ g \hA_\mu^p \times,   \nn\\
&\X_\mu\times \X_\nu=\sum_{p,q} \W_\mu^p 
\times \W_\nu^q,  \nn
\eea
so that 
\bea
&\vec{F}_{\mu\nu}=\hF_\mn + \hD _\mu \X_\nu 
-\hD_\nu \X_\mu + g\X_\mu \times \X_\nu  \nn\\
&=\sum_p \big[\dfrac23 \hF_\mn^p
+ (\hD_\mu^p \W_\nu^p-\hD_\mu^p \W_\nu^p) \big]  \nn\\
&+\sum_{p,q}\W_\mu^p \times \W_\nu^q.
\eea
With this we have the following form of SU(3) ECD \cite{ijmpa14}
\begin{gather}
{\cal L}_{ECD}= -\dfrac14 \F_\mn^2
=\sum_p \Big\{-\dfrac{1}{6} (\hF_\mn^p)^2 \nn\\
-\dfrac14 (\hD_\mu^p \W_\nu^p- \hD_\nu^p \W_\mu^p)^2 
-\dfrac{g}{2} \hF_\mn^p \cdot (\W_\mu^p \times \W_\nu^p) \Big\} \nn\\
-\sum_{p,q} \dfrac{g^2}{4} (\W_\mu^p \times \W_\mu^q)^2 \nn\\
-\sum_{p,q,r} \dfrac{g}2 (\hD_\mu^p \W_\nu^p
- \hD_\nu^p \W_\mu^p) \cdot (\W_\mu^q \times \W_\mu^r)  \nn\\
-\sum_{p\ne q} \dfrac{g^2}{4} \Big[(\W_\mu^p \times \W_\nu^q)
\cdot (\W_\mu^q \times \W_\nu^p)  \nn\\
+(\W_\mu^p \times \W_\nu^p)\cdot (\W_\mu^q \times \W_\nu^q) \Big],
\label{3ecd}
\end{gather}
which puts QCD to a totally different expression. 

We can add quarks in the Abelian decomposition,
\bea
&{\cal L}_{q} =\bar \Psi (i\gamma^\mu D_\mu-m) \Psi \nn\\
&= \bar \Psi (i\gamma^\mu \hD_\mu-m) \Psi 
+\dfrac{g}{2} \X_\mu \cdot \bar \Psi (\gamma^\mu \vec t) \Psi  \nn\\
&=\sum_p \Big[\bar \Psi^p (i\gamma^\mu \hD_\mu^p-m) \Psi^p
+\dfrac{g}{2} \W_\mu^p \cdot \bar \Psi^p
(\gamma^\mu \vec \tau^p) \Psi^p \Big], \nn\\
&\hD_\mu = \pro_\mu +\dfrac{g}{2i} {\vec t}\cdot \hA_\mu,
~\hD_\mu^p=\pro_\mu
+\dfrac{g}{2i} {\vec \tau^p}\cdot \hA_\mu^p,
\label{qlag}
\eea
where $m$ is the mass, $p$ denotes the color of the quarks, 
and $\Psi^p$ represents the three SU(2) quark doublets 
(i.e., $(r,b)$, $(b,g)$, and $(g,r)$ doublets) of the $(r,b,g)$ 
quark triplet. Notice that here we have suppressed the flavour 
degrees. 

\begin{figure}
\includegraphics[height=4cm, width=7cm]{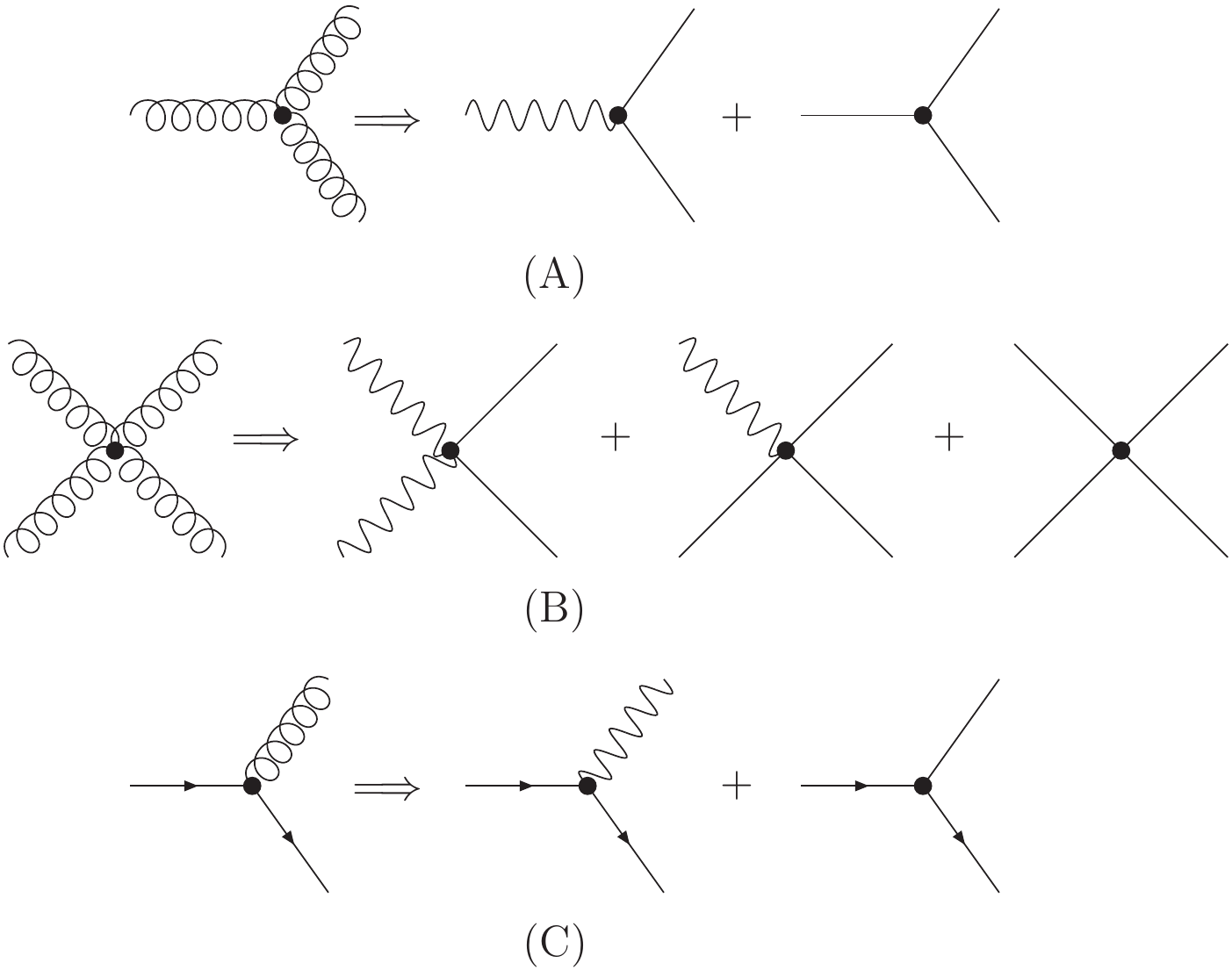}
\caption{\label{3ecdint} The decomposition of the Feynman 
diagrams in SU(3) QCD. The three-point and four-point gluon 
vertices are decomposed in (A) and (B), and the quark-gluon 
vertices are decomposed in (C). Notice the difference between 
SU(2) QCD and SU(3) QCD.}
\end{figure}

As we have emphasized, ECD does not change QCD. However, 
it reveals the important hidden characteristics of QCD and 
make them clearly visible. In the perturbative regime it refines 
the Feynman diagrams drastically. The decomposition of 
the Feynman diagram of the SU(3) QCD is shown graphically 
in Fig. \ref{3ecdint}. In (A) the three-point QCD gluon vertex 
is decomposed to two vertices, the one made of one neuron 
and two chromons and the other made of three chromons. 
In (B) the four-point gluon vertex is decomposed to three 
vertices made of one neuron and three chromons, two neurons 
and two chromons, and four chromons. In (C) the quark-gluon 
vertex is decomposed to the quark-neuron vertex and 
quark-chromon vertex. Obviously this decomposition of 
the Feynman diagrams is not possible in the conventional QCD 
where all gluons are treated equally.  

\begin{figure}
\includegraphics[height=5.5cm, width=6cm]{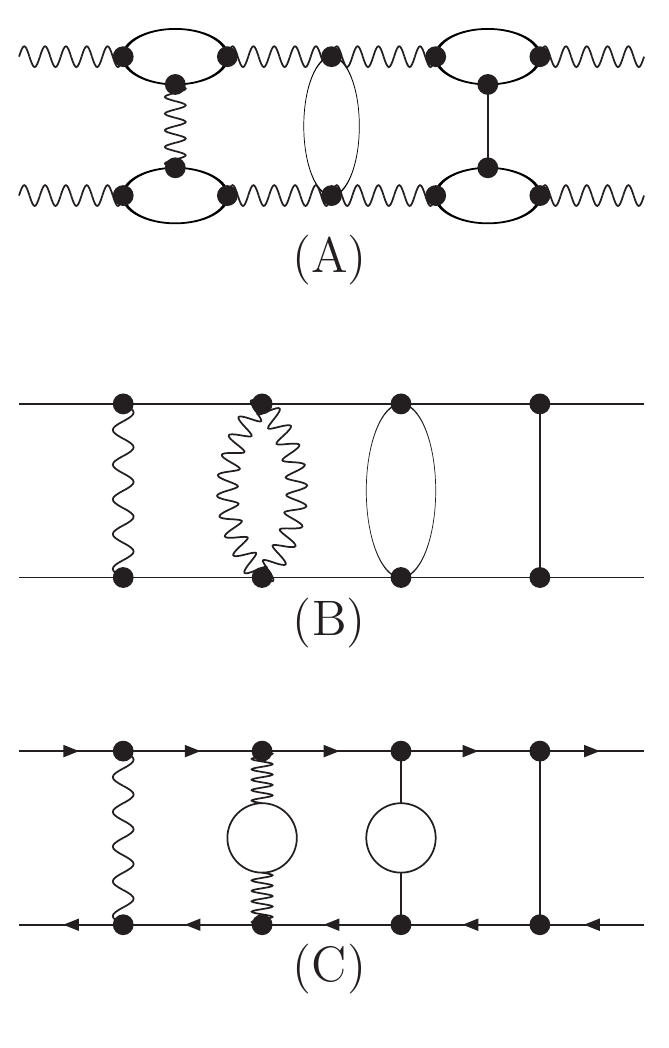}
\caption{\label{bind} The possible Feynman diagrams 
of the neuron and chromon interactions. Two neuron 
interaction is shown in (A), two chromon interaction is 
shown in (B), and quark-antiquark interaction is shown in (C).}
\end{figure}

There are two points to be emphasized here. First, the diagrams
do not contain the monopole. As we have already pointed out, 
this is because $\n$ represents the topological degree. Another 
reason why the monopole does not appear in the Feynman 
diagram is that the monopole makes the condensation, so that 
it disappears after the confinement sets in. So in the perturbative 
regime (inside the hadrons) only the neurons and chromons 
contribute to the Feynman diagrams. 

Second, the conservation of the color is explicit in the diagrams.
For example, the three-point vertex made of three neurons 
or two neurons and one chromon, and the four-point vertex 
made of three or four neurons are forbidden by the conservation 
of color. Moreover, the quark-neuron interaction does not 
change the quark color, but the quark-chromon interaction 
changes the quark color. 

This decomposition of the Feynman diagram should play 
important roles in the perturbative QCD. With this we can 
pinpoint what diagrams are actually responsible for the coupling 
constant renormalization and asymptotic freedom. Moreover, 
it allows us to demonstrate graphically that the neurons and 
chromons play totally different roles. To see this consider 
the three Feynman diagrams which describe the interaction 
of two neurons, two chromons, and quark-antiquark, shown 
in Fig. \ref{bind}. Clearly the neuron interaction in (A) looks 
exactly like the two photon interaction in QED. This is because 
the neurons are color neutral, so that they behave like 
the photons in QED. On the other hand, the chromon 
interaction in (B) looks exactly like the quark interaction 
shown in (C). Again this is because the chromons behave 
as colored source. The contrast between the neuron and 
chromon interactions is unmistakable. 

This has a deep implication in hadron spectroscopy. This 
means that the chromons can become the constituent of 
hadrons, so that they, just like the quarks, can form hadronic 
bound states which can be identified as the glueballs. 
Moreover, together with quarks they could form new hybrid 
baryons. But the neuron binding shown in (A) strongly implies 
that they can hardly make a bound state, which suggests that 
the neurons may not become the constituent of hadrons. 
This leads us to generalize the quark model to the quark 
and chromon model, which could provide a new picture 
of hadrons \cite{prd15,prd18}. 

This shows that the Abelian decomposition is not just 
a mathematical proposition. It can be tested directly by 
experiment. Since the neurons behave like the photons 
but the chromons behave like quarks, in the perturbative 
regime (i.e., in short distance) the neuron jet and the 
chromon jet should behave differently. So we could 
distinguish them and prove the existence of two types 
of gluon jets experimentally \cite{prd15,prd18}. 

In fact, we already have enough knowledge on how 
to differentiate the gluon jet from the quark jet 
experimentally \cite{jet1,jet2}. Moreover, 
we have a new proposal on how to separate different 
types of jets at LHC \cite{jet3}. Using these knowledges 
we could actually confirm the existence of two types of 
gluon jets experimentally. The confirmation of the gluon 
jet has justifed the asymptotic freedom \cite{wil}. 
The experimental confirmation of two types of gluon jets 
would be at least as important. 

\begin{figure}
\psfig{figure=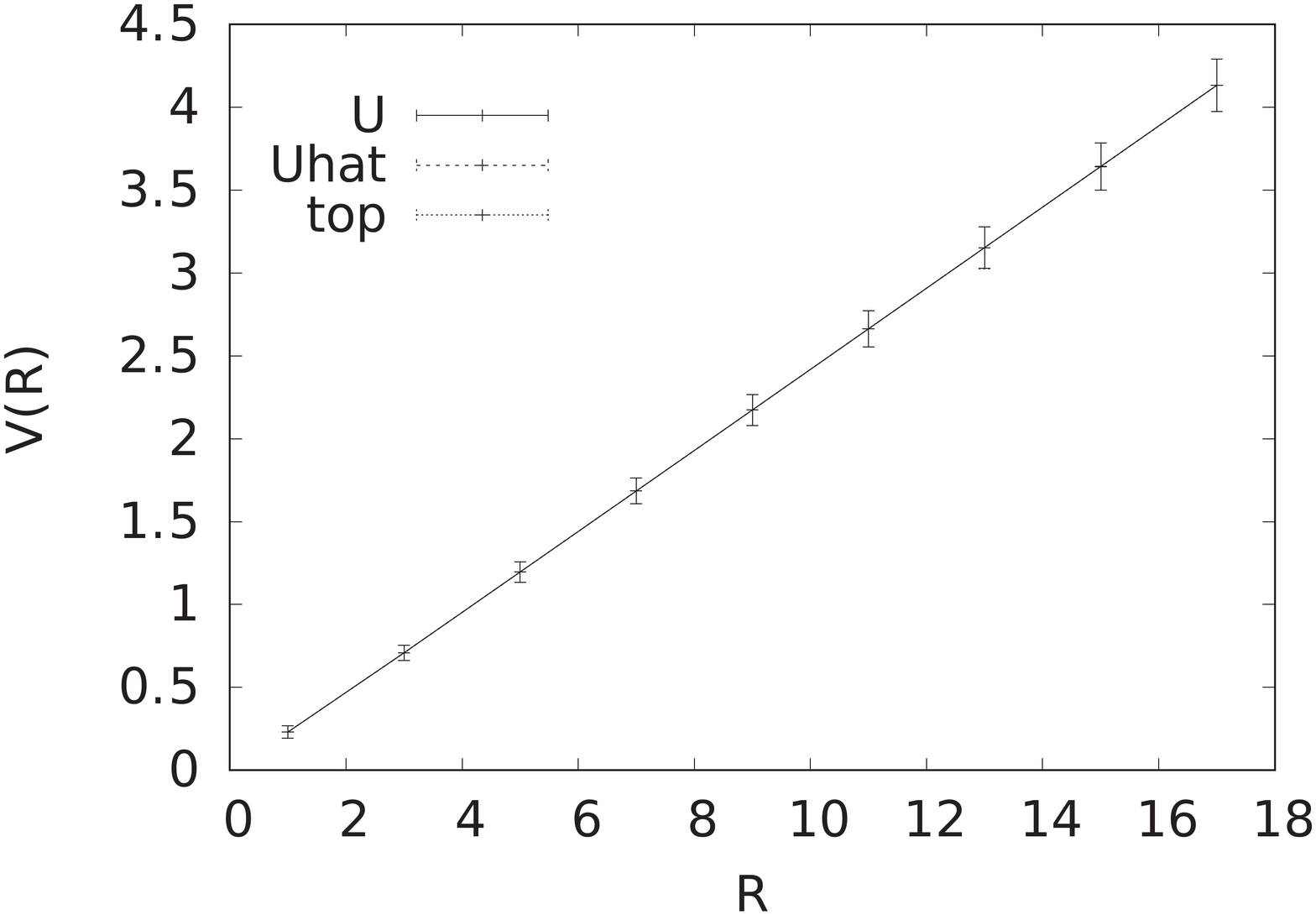, height=5cm, width=8cm}
\caption{\label{cundyn} The SU(3) lattice QCD calculation 
which establishes the monopole dominance in the confining 
force in Wilson loop. Here the confining forces shown in full, 
dashed, and dotted lines are obtained with the full potential, 
the Abelian potential, and the monopole potential, 
respectively.}
\end{figure}

But the Abelian decomposition plays the decisive role in 
the non-perturbative regime. First of all, it allows us to 
prove not only the Abelian dominance but also the monopole 
dominance in QCD rigorously. Obviously the chromons can 
not play any role in the confinement, because they themselves 
have to be confined. This can be provided theoretically. First, 
we can show that only the restricted potential contributes to 
the Wilson loop integral which generates the linear confining 
potential \cite{prd02}. This is the Abelian dominance. 

Moreover, we can argue that actually the monopole part of 
the restricted potential is responsible for the confining 
potential. This is because the Maxwell part plays the role 
of the electromagnetic potential in QED, which is known 
to have no confinement. This demonstrates the monopole 
dominance, that the monopole is responsible for 
the confinement.      

This is backed up numerically in the lattice QCD. Implementing 
the Abelian decomposition on the lattice, we can show 
that the confining force comes from the monopole part 
of the restricted potential \cite{kondo,cundy}. The recent 
result of SU(3) lattice calculation is shown in Fig. \ref{cundyn}, 
which clearly tells that all three potentials, the full potential, 
the Abelian potential, and the monopole potential generate 
the same confining force. This establishes the monopole 
dominance.

The monopole dominance, however, does not tell us how 
the monopole confines the color. Fortunately, the Abelian 
decomposition allows us to show how. To see this it is 
important to understand that field theoretically ECD puts QCD 
in the background field formalism \cite{prd01,dewitt,pesk}. 
This is because in ECD the restricted potential and the valence 
potential can be treated as the slow varying classical field and 
the fluctuating quantum field. 

This enlarges the gauge symmetry of QCD and makes 
ECD to have two independent color gauge symmetries, 
the classical and quantum gauge symmetry. This has 
a deep consequence. For example, the quantum gauge 
symmetry keeps the chromons massless, even though 
they transform gauge covariantly \cite{prd01}.     

More importantly, the background field formalism provides 
us an ideal platform to calculate the QCD effective action 
gauge independently. This is because we can treat the slow 
varying classical part as the background, and integrate 
the quantum part to obtain the one-loop effective action. 

Furthermore, ECD simplifies the complicated non-Abelian 
gauge symmetry to a simple discrete symmetry called 
the color reflection symmetry which is much easier to 
handle \cite{prd80,prl81}. This plays a crucial role  
for us to implement the gauge invariance when we 
calculate the QCD effective action to demonstrate 
the monopole condensation, and allows us to clarify 
the confusions of the old calculations of the effective 
action \cite{ prd13,ijmpa14}. 
 
But most importantly the Abelian decomposition allows us 
to calculate the SU(3) QCD effective action directly from 
the SU(2) QCD effective action \cite{ijmpa14}. This is because 
the Abelian decomposition of SU(3) QCD transforms it to 
a Weyl symmetric form of three SU(2) QCD. This is evident    
in (\ref{3ecd}) and (\ref{qlag}), which tell that SU(3) QCD is 
Weyl symmetric, symmetric under the permutation of three 
SU(2) subgroups. As we will see, this Weyl symmetry greatly 
simplifies the calculation of the SU(3) QCD effective action. 
In general we can show that the Weyl symmetric Abelian 
decomposition of SU(N) QCD allows us to obtain the SU(N) 
QCD effective action directly from the SU(2) QCD effective 
action. 

\section{Color Reflection Invariance}

As we have emphasized, the Abelian decomposition is gauge 
independent. On the other hand, the selection of the Abelian 
direction amounts to the gauge fixing. So, once we fix 
the Abelian direction the gauge symmetry is broken. On 
the other hand, this  does not break the gauge symmetry 
completely, so that we have a residual discrete symmetry 
called the color reflection symmetry even after the Abelian 
decomposition \cite{prd80,prl81,prd13}.

The importance of this residual symmetry comes from 
the following observation. First, this plays the role of 
the gauge symmetry after the Abelian decomposition. 
Second, this symmetry is much simpler than the color 
gauge symmetry. This tells that the Abelian decomposition
reduces the complicated non-Abelian gauge symmetry to 
a simple discrete symmetry which is much easier to 
handle. This greatly helps us to implement the gauge 
invariance in the calculation of the QCD effective action. 
So we discuss the color reflection symmetry first. 

Consider the SU(2) QCD first and make the color 
reflection, the $\pi$-rotation of the SU(2) basis along 
the $\n_2$-direction which inverts the color direction 
$\n$,  
\bea
(\n_1,\n_2,\n) \rightarrow (-\n_1,\n_2,-\n).
\label{2cref}
\eea 
Obviously this is a gauge transformation which should 
not change the physics. On the other hand, under the color 
reflection (\ref{2cref}) we have \cite{prd13}
\begin{gather}
\hA_\mu \rightarrow \hA_\mu^{(c)}
= -A_\mu \n-\oneg \n \times \pro_\mu \n,  \nn\\
A_\mu \rightarrow A_\mu^{(c)}= -\n \cdot \A_\mu=-A_\mu.  
\label{2ncrt}
\end{gather}
Moreover, 
\begin{gather}
\X_\mu \rightarrow \X_\mu^{(c)}
=-(X_\mu^1~\n_1-X_\mu^2~\n_2),  \nn
\end{gather}
or, in the complex notation by
\begin{gather}
R_\mu \rightarrow R_\mu^{(c)}=-\bR_\mu,
\label{2ccrt}
\end{gather}
where $\bR_\mu=(X_\mu-iX_\mu^2)/\sqrt 2$. 

But since the isometry condition (\ref{ap}) is insensitive to (\ref{2cref}), we have two different Abelian decompositions 
imposing the same isometry, 
\begin{gather}
\A_\mu= \hA_\mu+\X_\mu,
~~~~\A_\mu= \hA_\mu^{(c)}+\X_\mu^{(c)}, 
\end{gather}
without changing the physics. This is why the color 
reflection (\ref{2cref}) becomes a discrete symmetry of 
QCD after the Abelian decomposition \cite{prd80,prl81}. 

To understand the meaning of this, notice that the neuron 
potential $A_\mu$ change the signature, while the topological 
part remains invariant. Moreover the chromon changes to 
the complex conjugate partner (together with the change of 
the signature), which changes the chromon to anti-chromon 
and flips the sign of the chromon charge.  

This is not surprising. In the absence of the topological part 
(\ref{2ecd}) describes QED which is coupled to the massless 
charged vector field where the neuron plays the role of 
the photon. And in QED it is well known that the photon has 
negative charge conjugation quantum number. So it is natural 
that $A_\mu$ in SU(2) QCD changes the signature under 
the color reflection. Similarly we can argue that $A_\mu$ 
changes the signature under the parity \cite{prd13}.   

On the other hand the monopole potential remains unchanged 
under the color reflection. This means that the monopole 
and anti-monopole are physically undistinguishable in 
QCD \cite{prd80,plb82}. This should be contrasted with 
the monopole in spontaneously broken gauge theories, where 
the monopole and ant-monopole are physically different.  

This confirms that, although there are two possible 
magnetic backgrounds, only the monopole background 
coming from $\cC_\mu$ is qualified to be the legitimate 
background we can choose in the calculation of the QCD 
effective action. This is because $A_\mu$ changes 
the signature under the color reflection and thus fails 
to be gauge invariant. Indeed this is the reason why 
the Savvidy vacuum is not gauge invariant. 

As importantly, (\ref{2ccrt}) tells that the physics should 
not change when we change the chromon to anti-chromon. 
In fact (\ref{2ccrt}) tells that the chromon and anti-chromon 
are the color reflection partner. This means that they can 
not be separately discussed in QCD and should always play 
exactly the same amount of role. This is the reason why 
the color should become unphysical and confined, which 
makes QCD totally different from QCD. This point plays 
a crucial role when we implement the gauge invariance in 
the calculation of the effective action \cite{prd80,prl81,prd13}

In the fundamental representation the color reflection 
(\ref{2cref}) is given by the 4 element subgroup of SU(2)
made of \cite{prd80,prl81}
\bea
&C_1=\left(\begin{array}{cc}
1 & 0 \\ 0 & 1 \\ \end{array} \right),
~~C_2=\left(\begin{array}{cc}
-1 & 0 \\ 0 & -1 \\ \end{array} \right), \nn\\
&C_3=\left(\begin{array}{cc}
0 & 1 \\ -1 & 0 \\ \end{array} \right),
~~C_4=\left(\begin{array}{cc}
0 & -1 \\ 1 & 0 \\ \end{array} \right).
\label{crg2}
\eea
This can be expressed by
\begin{gather}
C_k=D_a R_b, ~~(a=1,2;~b=1,2;~k=1,2,...,4),  \nn\\
D_1=\left(\begin{array}{cc}
1 & 0 \\ 0 & 1 \\ \end{array} \right),
~~~D_2=\left(\begin{array}{cc}
-1 & 0 \\ 0 & -1 \\ \end{array} \right)  \nn\\
R_1=\left(\begin{array}{cc}
1 & 0 \\ 0 & 1 \\ \end{array} \right),
~~~~R_2=\left(\begin{array}{cc}
0 & 1 \\ -1 & 0 \\ \end{array} \right),
\end{gather}
which contains the diagonal subgroup made of $D_1$ and 
$D_2$. This becomes the residual symmetry of the SU(2) 
quark doublet $(r,b)$ after the Abelian decomposition. 
Notice that $R_2$ plays the role of the generator of the color 
reflection group.

As for the gluons which form the adjoint representation 
the color reflection can be simplified further for the following 
reasons. First, the diagonal subgroup has no effect on 
the adjoint representation. Second, the color reflection 
changes $\n$ to $-\n$ and $(\n_1,\n_2)$ to $(-\n_1,\n_2)$. 
So, the gluon triplet is decomposed to two independent 
representations.

Indeed, for 
the neuron we have 
\begin{gather}
R_2:~~A_\mu \rightarrow -A_\mu. 
\label{2nrep}
\end{gather}
But for the chromon we have
\begin{gather}
R_2:~~(\X_\mu,\X_\mu^{(c)}) 
\rightarrow -(\X_\mu^{(c)},\X_\mu), \nn
\end{gather}
or equivalently
\begin{gather}
R_2:~~(R_\mu,\bR_\mu) \rightarrow -(\bR_\mu,R_\mu).
\label{2crep}
\end{gather}
This confirms that the neuron and chromon transform 
independently, forming one-dimensional and 
two-dimensional representations under the color 
reflection. This drastically simplifies the non-Abelian 
gauge symmetry. 

For SU(3) the fundamental representation the color reflection 
group is made of 24 elements subgroup of SU(3) given 
by \cite{prd80,prl81,kondor}
\begin{gather}
C_k=D_a R_b,  \nn\\
(a=1,2,3,4;~b=1,2,...,6;~k=1,2,...,24),   \nn\\
D_1=\left(\begin{array}{ccc}
1 & 0 & 0 \\ 0 & 1 & 0 \\ 0 & 0 & 1 \\
\end{array} \right),
~~~D_2=\left(\begin{array}{ccc}
-1 & 0  & 0 \\ 0 & -1 & 0 \\ 0 & 0 & 1 \\
\end{array} \right),  \nn\\
D_3=\left(\begin{array}{ccc}
1 & 0  & 0 \\ 0 & -1 & 0 \\ 0 & 0 & -1 \\
\end{array} \right), 
D_4=\left(\begin{array}{ccc}
-1 & 0 & 0 \\ 0  & 1 & 0 \\ 0 & 0 & -1 \\
\end{array} \right),  \nn\\
R_1=\left(\begin{array}{ccc}
1 & 0 & 0 \\ 0 & 1 & 0 \\ 0 & 0 & 1 \\
\end{array} \right),
~~~~R_2=\left(\begin{array}{ccc}
0  & 1 & 0 \\ -1 & 0 & 0 \\ 0  & 0 & 1 \\
\end{array} \right), \nn\\
R_3=\left(\begin{array}{ccc}
1 & 0 & 0 \\ 0 & 0 & 1 \\ 0 & -1 & 0 \\
\end{array} \right),
~~~R_4=\left(\begin{array}{ccc}
0 & 0 & 1 \\ 0 & -1 & 0 \\ 1 & 0 & 0 \\
\end{array} \right), \nn\\
R_5=\left(\begin{array}{ccc}
0 & 1 & 0 \\ 0 & 0 & 1 \\ 1 & 0 & 0 \\
\end{array} \right),
~R_6=\left(\begin{array}{ccc}
0  & 0 & 1 \\ -1 & 0 & 0 \\ 0 & -1 & 0 \\
\end{array} \right),
\label{crg3}
\end{gather}
where the four $D$-matrices form the diagonal subgroup. 
This describes the residual symmetry of the quark triplet 
$(r,b,g)$ after the Abelian decomposition. Notice that here 
$R_2$ and $R_3$ play the role of the generator. For example, 
we have $R_5=R_3\cdot R_2$, $R_6=R_2\cdot R_3$, and 
$R_4=R_2\cdot R_3\cdot R_2$.

For the gluon octet which form the adjoint representation of 
SU(3) the color reflection can be simplified further. Just as in 
SU(2) QCD, the neurons and chromons transform separately,
among themselves. To see exactly how they transform 
notice that, according to (\ref{cp3}) and (\ref{cdec3}) 
the two neurons form a (mutually dependent) triplet 
$(A_\mu^1,A_\mu^2,A_\mu^3)$ and the six chromons form 
a sextet $(\R_\mu,\B_\mu,\G_\mu,\R_\mu^{(c)},\B_\mu^{(c)},
\G_\mu^{(c)})$ or equivalently 
$(R_\mu,B_\mu,G_\mu,\bR_\mu,\bB_\mu,\bG_\mu)$. For 
the neurons we have
\begin{gather}
R_2:~~~~\left(\begin{array}{c} A_\mu \\ A_\mu' \end{array} \right)
\rightarrow \left(\begin{array}{cc}
-1 & 0 \\ 0 & 1 \\ \end{array} \right)  
\left(\begin{array}{c} A_\mu \\ A_\mu' \end{array} \right), \nn\\
R_3:\left(\begin{array}{c} A_\mu \\ A_\mu' \end{array} \right)
\rightarrow \left(\begin{array}{cc}
1/2 & \sqrt 3/2 \\ \sqrt 3/2 & -1/2 \\ 
\end{array} \right)
\left(\begin{array}{c} A_\mu \\ A_\mu' \end{array} \right),
\end{gather}
from which we have
\begin{gather}
R_2:~~(A_\mu^1,A_\mu^2,A_\mu^3) 
\rightarrow -(A_\mu^1,A_\mu^3,A_\mu^2),  \nn\\
R_3:~~(A_\mu^1,A_\mu^2,A_\mu^3) 
\rightarrow -(A_\mu^3,A_\mu^2,A_\mu^1),  \nn\\
R_4:~~(A_\mu^1,A_\mu^2,A_\mu^3) 
\rightarrow -(A_\mu^2,A_\mu^1,A_\mu^3),  \nn\\
R_5:~~(A_\mu^1,A_\mu^2,A_\mu^3) 
\rightarrow ~~(A_\mu^3,A_\mu^1,A_\mu^2),  \nn\\
R_6:~~(A_\mu^1,A_\mu^2,A_\mu^3) 
\rightarrow ~~(A_\mu^2,A_\mu^3,A_\mu^1).
\label{3ncrt}
\end{gather}
This tells that basically $R_2,~R_3,~R_4$ represent 
the permutations of two SU(2) neurons (up to the signature 
change), but $R_5,~R_6$ represent the cyclic permutations 
of three SU(2) neurons.
 
For the chromons the color reflection acts as follows,
\begin{gather}
R_2:~~(R_\mu,B_\mu,G_\mu,\bR_\mu,\bB_\mu,\bG_\mu)  \nn\\
~~~~~~~~\longrightarrow (\bR_\mu,\bG_\mu,\bB_\mu,R_\mu,G_\mu,B_\mu),  \nn\\
R_3:~~(R_\mu,B_\mu,G_\mu,\bR_\mu,\bB_\mu,\bG_\mu)  \nn\\
~~~~~~~~\longrightarrow -(\bG_\mu,\bB_\mu,\bR_\mu,G_\mu,B_\mu,R_\mu),  \nn\\
R_4:~~(R_\mu,B_\mu,G_\mu,\bR_\mu,\bB_\mu,\bG_\mu)  \nn\\
~~~~~~~~\longrightarrow -(\bB_\mu,\bR_\mu,\bG_\mu,B_\mu,R_\mu,G_\mu),     \nn\\
R_5:~~(R_\mu,B_\mu,G_\mu,\bR_\mu,\bB_\mu,\bG_\mu)  \nn\\
~~~~~~~~\longrightarrow -(G_\mu,R_\mu,B_\mu,\bG_\mu,\bR_\mu,\bB_\mu),  \nn\\
R_6:~~(R_\mu,B_\mu,G_\mu,\bR_\mu,\bB_\mu,\bG_\mu)  \nn\\
~~~~~~~~\longrightarrow -(B_\mu,G_\mu,R_\mu,\bB_\mu,\bG_\mu,\bR_\mu).
\label{3ccrt}
\end{gather} 
Here $R_2,~R_3,~R_4$ denote the anti-chromon transformation 
(complex conjugation) plus permutations of two chromons, but 
$R_5,~R_6$ denote the cyclic permutations of three chromons 
(up to the signature change). Just as in SU(2) QCD here 
the complex conjugation (anti-chromon transformation) of 
the chromons in the color reflection plays the crucial role in 
the calculation of the effective action.  

The above analysis reveals another important difference 
between the neuron and chromon. Clearly (\ref{3ncrt}) tells 
that the neurons  permute amomg themselves, but (\ref{3ccrt}) 
tells that the chromons transform to anti-chromons, under 
the color reflection. In other words, just like the photon in 
QED the neurons have no anti-neurons. In comparison 
the chromons have the anti-chromon partners. This is 
because the neurons are color neutral so that they have 
the real representation, while the chromons are colored 
and allow the complex representation.

At this point one might wonder if there is any relation between 
the color reflection group and Weyl group. For SU(3), the Weyl 
group is the six elements permutation group of three colors 
which has a three-dimensional representation given by
\begin{gather}
W_1=\left(\begin{array}{ccc}
1 & 0 & 0 \\ 0 & 1 & 0 \\ 0 & 0 & 1 \\
\end{array} \right),
~~~W_2=\left(\begin{array}{ccc}
0  & 1 & 0 \\ 1 & 0 & 0 \\ 0  & 0 & 1 \\
\end{array} \right), \nn\\
W_3=\left(\begin{array}{ccc}
1 & 0 & 0 \\ 0 & 0 & 1 \\ 0 & 1 & 0 \\
\end{array} \right),
~~~W_4=\left(\begin{array}{ccc}
0 & 0 & 1 \\ 0 & 1 & 0 \\ 1 & 0 & 0 \\
\end{array} \right), \nn\\
W_5=\left(\begin{array}{ccc}
0 & 1 & 0 \\ 0 & 0 & 1 \\1 & 0 & 0 \\
\end{array} \right),  
~~W_6=\left(\begin{array}{ccc}
0  & 0 & 1 \\ 1 & 0 & 0 \\ 0 & 1 & 0 \\
\end{array} \right), 
\label{wg3}
\end{gather}
which contains the cyclic $Z_3$ made of $W_1,~W_5,$ and 
$W_6$.

This tells that the two groups are different. They have different 
origin. The Weyl group comes as the symmetry of the Abelian 
decomposition, but the color reflection group comes as the residual 
symmetry of the Abelian decomposition. Unlike the color 
reflection group (\ref{crg3}), the Weyl group (\ref{wg3}) is 
not a subgroup of SU(3). Moreover, the Weyl group has no 
complex conjugation operation which transforms the chromons 
to anti-chromons. On the other hand they have a common 
subgroup $Z_3$, the cyclic permutation group of three colors.  

Both the color reflection group and the Weyl group play 
a fundamental role in hadron spectroscopy. Only the color 
reflection invariant and Weyl invariant combinations of 
quarks and gluons can become physical in the quark and 
chromon model \cite{prd15}. Moreover, they play the crucial 
role for us to calculate the one-loop effective action of SU(3)
QCD and prove the monopole condensation gauge invariantly.

\section{One-Loop Effective Action of SU(2) QCD: A Review}

Before we calculate the one-loop effective action of SU(3) 
QCD we need to understand how we calculate the effective 
action of the SU(2) QCD, for two reasons.  First, the calculation 
of the SU(2) QCD effective action becomes an essential part 
for the calculation of the effective action of SU(3) QCD. Second, 
the early calculations had critical defects \cite{savv,niel,ditt,yil}. 
The Savvidy vacuum was unstable. More seriously, it was not 
gauge invariant. So we have to know how to correct 
these critical mistakes.

To obtain the one-loop effective action we must divide 
the gauge potential to the classical and quantum parts 
and integrate out the quantum part in the presence of 
the classical background. The Abelian decomposition 
naturally provides an ideal platform for this, since we 
can treat the Abelian part as the classical background 
and integrate out the valence part. 

Imposing the quantum gauge fixing condition  
$\bD_\mu \X_\mu=0$ we have \cite{prd02,jhep05},
\begin{gather}
\exp \big[iS_{eff}(\hA_\mu)\big]
=\Int \cD \X_\mu \cD \X_\mu^{(c)} \cD \vec c \cD {\vec c}^* \nn\\
\exp \Big\{i\Int \big[-\dfrac{1}{4} \hF^2_\mn
-\dfrac{1}{4}(\hD_\mu\X_\nu-\hD_\nu\X_\mu)^2 \nn\\
-\dfrac{g}{2} \hF_\mn\cdot (\X_\mu \times \X_\nu)
-\dfrac{g^2}{4} (\X_\mu \times \X_\nu)^2  \nn\\
+{\vec c}^* \bD_\mu D_\mu \vec c 
-\dfrac{1}{2\xi} (\bD_\mu \X_\mu)^2 \big] d^4 x \Big\},
\label{2ea0}
\end{gather}
where $\vec c$ and ${\vec c}^*$ are the ghost fields. But notice 
that the quartic interaction of $\X_\mu$ can be neglected since 
this does not contribute in the one-loop integration.

The above integral expression of the effective action has 
the following advantages. First, the separation of the classical 
and quantum parts is explicitly gauge independent. Second, 
the functional integration of is made by the chromon and 
anti-chromon in the color reflection symmetric way. As we 
will see, this point plays the crucial role for us to implement 
the gauge invariance in the functional integral. 

Clearly these salient features were lacking in the old 
calculations \cite{savv,niel,ditt,yil}. The classical part 
was not Abelian, and the separation of the classical and 
quantum parts was ad hoc. Moreover, all three gluons 
were integrated in the functional integration. In particular, 
the role of the anti-chromon in the functional integral
was completely obscure, which has made the implementation 
of the gauge invariance very difficult in the functional integral. 
This was because the Abelian decomposition was not 
available at that time.   

A more serious problem of the old calculations, however, was 
the wrong background. The Abelian decomposition tells that 
there are two possible backgrounds, the non-topological 
$F_\mn$ and the topological $H_\mn$. But in old calculations 
people have chosen $F_\mn$ which is not gauge invariant 
nor parity conserving. Moreover, this does not describe 
the monopole background. This is the problem with 
the Savvidy vacuum \cite{savv,niel,ditt,yil}. 

Let us choose the wrong background for the moment, and let
\bea
&\bar F_\mn= H \delta_{[\mu}^1 \delta_{\nu]}^2.
\label{savb}
\eea
where $H$ is a constant magnetic field of $F_\mn$ in 
$z$-direction. With this the effective action is expressed 
by the chromon and ghost loop determinants given by $K$ 
and $M$ \cite{savv,niel,ditt,yil},
\bea
&\Delta S = \dfrac{i}{2} \ln {\rm Det} K 
- i \ln {\rm Det} M,  \nn\\
&{\rm Det}^{-1/2} K_\mn ={\rm Det} \Big(-g_\mn \bD^2
+2ig \bar F_\mn \Big),\nn \\
& {\rm Det} M^{1/2} = {\rm Det} \big(-\bD^2 \big),
\label{2eax}
\eea
where $\bar D_\mu=\pd_\mu-g \bar A_\mu \n \times$ is 
the covariant derivative defined by the classical background.

\begin{figure}
\psfig{figure=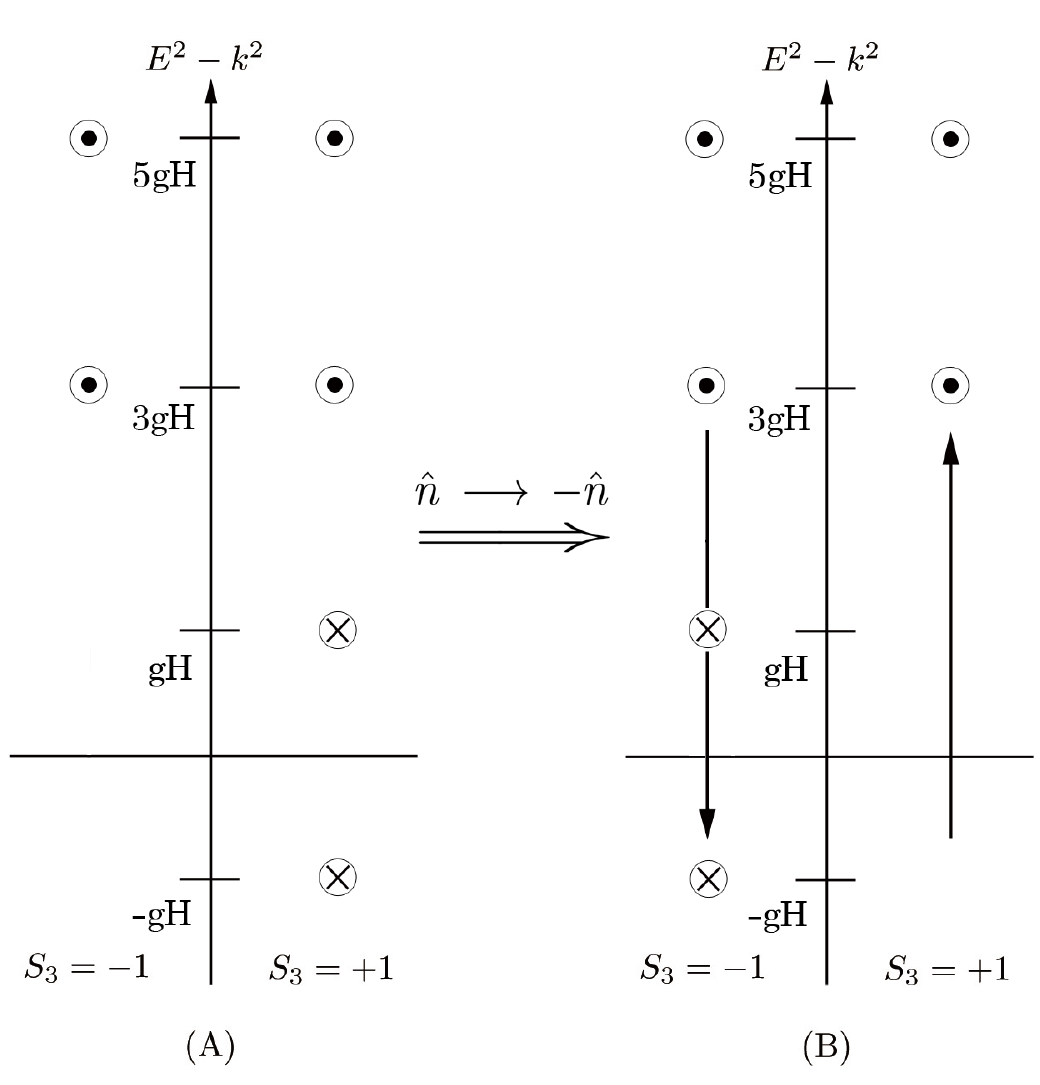, height=4cm, width=6cm}
\caption{\label{evsb2} The gauge invariant eigenvalues 
of the gluon functional determinant. Notice that the color 
reflection invariance removes the tachyonic modes in both
(A) and (B).}
\end{figure}

One can calculate the functional determinant of the gluon
loop from the energy spectrum of a massless charged vector 
field moving around the constant magnetic field $H$, which 
is given by \cite{tsai}
\bea
&E^2 = 2g H (n + \dfrac{1}{2} - q S_3) + k^2,     
\label{ev}
\eea
where $S_3$ and $k$ are the spin and momentum of 
the vector field in the direction of the magnetic field, 
and $q=\pm 1$ is the charge of the vector field. Notice 
that the energy spectrum of gluons for two different 
spin polarizations $S_3=\pm 1$ is different. Moreover, 
when $S_3=1$, it contains negative (tachyonic) eigenvalues 
which violate the causality. This is schematically shown 
in Fig. \ref{evsb2} for $q=+1$ in (A) and $q=-1$in (B). 

From this one has the integral expression of the effective 
action given by \cite{savv,niel,ditt,yil}
\bea
&\Delta S = i \ln {\rm Det} [(-\bD^2+gH)(-\bD^2-gH)] \nn\\
&- 2i \ln {\rm Det}(-\bD^2),   \nn\\
&\Delta{\cal L} = \lim_{\epsilon \rightarrow 0}
\dfrac{\mu^2}{16 \pi^2}\Int_{0}^{\infty}
\dfrac{dt}{t^{2-\epsilon}} \dfrac{gH}{\sinh (gHt/\mu^2)} \nn\\
&\times \Big[\exp (-2gHt/\mu^2 )
+ \exp (+2gHt/\mu^2) \Big].
\label{2eahx}
\eea
Notice that the two exponential terms which represent 
the contribution of two spin polarizations is different. 
Moreover, the second term has a severe infra-red 
divergence. 

With the standard $\zeta$-function regularization one 
can integrate (\ref{2eahx}) and obtain the SNO effective 
action \cite{savv,niel,ditt}
\bea
&{\cal L}_{eff}=-\dfrac12 H^2-\dfrac{11g^2}{48\pi^2} H^2
(\ln \dfrac{gH}{\mu^2}-c) \nn\\
&+ i \dfrac {g^2} {8\pi} H^2,
\label{snoea}
\eea
where $c$ is an integration constant. This contains 
the well-known imaginary part coming from 
the tachyonic eigenstates, which destablizes 
the Savvidy vacuum \cite{niel,ditt,yil}. 

There have been huge efforts to cure this instability of 
the Savvidy vacuum \cite{niel,ditt,yil}. Actually there are 
ways to cure the instability. One way is to impose the causality 
in the functional integral \cite{prd02}.  Clearly the causality 
removes the tachyonic modes, and remove the imaginary part.
Another is to calculate the imaginary part perturbatively to 
the second order in the coupling constant $g$ \cite{jhep05}. 
This is because the imaginary part at one loop level is in 
the order $g^2$, although in principle the effective action 
is non-perturbative. And the perturbative calculation confirms 
that there should be no imaginary part.

But we emphasize that this instability is not the only problem 
of the Savvidy vacuum. There are other problems. For instance, 
it does not describe the monopole condensation. But the most 
serious problem is that it is not gauge invariant nor parity 
conserving, and thus can not be identified as the QCD vacuum.   

Another serious defect in the above calculation is that the gauge 
invariance is not correctly implemented in the calculation. We 
could implement the gauge invariance in the old calculations,
but there is no point to do so. First of all, this does not change 
the result. More seriously, it is meaningless and irrelevant to do 
so because the gauge invariance has already been compromised 
as soon as the non-topological background (\ref{savb}) was 
chosen. 
 
Fortunately the Abelian decomposition allows us to calculate 
the effective action correctly \cite{prd02,jhep05,prd13}. 
First, it allows us to separate the gauge invariant and parity 
conserving monopole background gauge independently. 
Second, it allows us to impose the much simpler color 
reflection invariance to implement the gauge invariance 
in the calculation of the effective action. This makes 
the calculation simple and clear.

To show this we first make the Abelian decomposition and 
choose the gauge invariant and parity conserving monopole 
background $H_\mn$ \cite{prd13,ijmpa14}
\bea
&\bH_\mn=\bH \delta_{[\mu}^1 \delta_{\nu]}^2, 
\label{mb}
\eea
where $\bH$ now is a constant chromo-magnetic field of 
$H_\mn$ in $z$-direction. With this we can integrate out 
the chromon and express the effective action by the chromon 
and ghost loop determinants \cite{prd02,jhep05,prd13},
\bea
&\Delta S = \dfrac{i}{2} \ln {\rm Det} K 
- i \ln {\rm Det} M,  \nn\\
&{\rm Det}^{-1/2} K_\mn ={\rm Det} \Big(-g_\mn \bD^2
+2ig \bH_\mn \Big),\nn \\
& {\rm Det} M^{1/2} = {\rm Det} \big(-\bD^2 \big),
\label{ea0}
\eea
where $\bar D_\mu$ is the covariant derivative defined by 
the monopole background. 

The next step is to calculate the chromon loop functional 
determinant implementing the gauge invariance. Clearly 
the energy spectrum of the chromon moving around 
the constant magnetic field $\bar H$ is given by (\ref{ev}) 
as before, but here we have to find the gauge invariant 
energy spectrum. A simplest way to do that is to choose 
the energy eigenvalues which remain invariant under 
the color reflection. This is because the color reflection 
invariance is synonymous to the gauge invariance after 
the Abelian decomposition. 

Under the color reflection the chromon undergoes to 
the complex conjugation and becomes anti-chromon which 
has opposite color charge. So the energy spectrum shown 
in Fig. \ref{evsb2} (A) for $q=+1$ changes to (B) for $q=-1$. 
As we have emphasized, however, physics should not change 
under this color reflection. In particular, the eigenvalues of 
the chromon functional determinant for each spin polarization 
should remain the same. 

This means that only the eigenvalues which appear in both (A) 
and (B) simultaneously become gauge invariant and physical. 
This excludes the lowest two (in particular the tachyonic) 
eigenvalues in both (A) and (B). This is the C-projection which 
removes the tachyonic modes and makes the monopole 
condensation stable \cite{prd13}. 

This tells that the color reflection invariance plays exactly 
the same role as the G-parity in string theory. In the NSR 
string theory the GSO projection restores the supersymmetry 
and modular invariance by projecting out the tachyonic 
vacuum \cite{gso,witt}. Just like the GSO-projection in string 
theory, the C-projection in QCD removes the tachyonic 
modes and restores the gauge invariance which assures 
the stable monopole condensation. 

So, imposing the color reflection invariance we 
have \cite{prd02,jhep05,prd13},
\bea
&\Delta S = i \ln {\rm Det} [(-\bD^2+g\bH)(-\bD^2+g \bH)] \nn\\
&- 2i \ln {\rm Det}(-\bD^2),   \nn\\
&\Delta{\cal L} = \lim_{\epsilon \rightarrow 0}
\dfrac{\mu^2}{16 \pi^2}\Int_{0}^{\infty}
\dfrac{dt}{t^{2-\epsilon}} \dfrac{g \bH}{\sinh (g \bH t/\mu^2)} \nn\\
&\times \Big[\exp (-2g \bH t/\mu^2 )
+ \exp (-2g \bH t/\mu^2) \Big].
\label{eaho}
\eea
This should be compared with (\ref{2eahx}). Here the two 
chromon spin polarization contributions shown in the two 
exponential terms become identical. Moreover, we have no 
infra-red divergence here. This, of course, is because the color reflection invariance removes the eigenvalues which are not 
gauge invariant. Integrating this we have the SU(2) QCD 
effective action
\bea
&{\cal L}_{eff}=-\dfrac12 \bH^2-\dfrac{11g^2}{48\pi^2} \bH^2
(\ln \dfrac{g \bH}{\mu^2}-c)
\label{2eah}
\eea
Clearly the effective action has no imaginary part at all. 

\begin{figure}[t]
\epsfig{figure=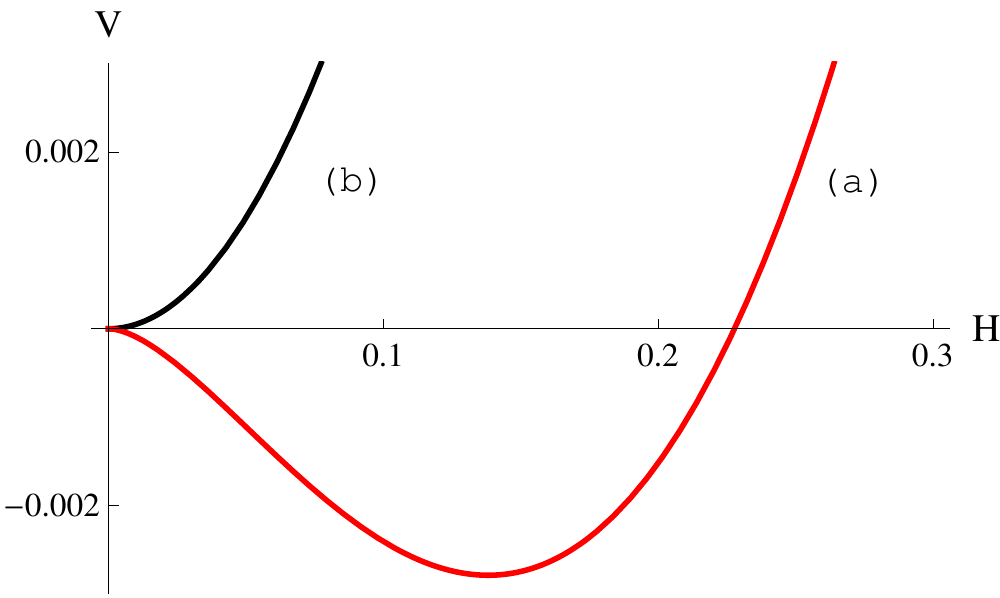, height = 3.5 cm, width = 6 cm}
\caption{\label{2pot} The effective potential of SU(2)
QCD in the pure magnetic background. Here (a) is the effective 
potential and (b) is the classical potential.}
\end{figure}

The effective action (\ref{2eah}) generates the much desired
dimensional transmutation in QCD. From (\ref{2eah}) we have 
the following effective potential 
\bea
V=\dfrac12 \bH^2
\Big[1+\dfrac{11 g^2}{24 \pi^2}(\ln\dfrac{g \bH}{\mu^2}-c)\Big].
\eea
With this we define the running coupling $\bar g$ by \cite{prd02,jhep05}
\begin{gather}
\dfrac{\pro^2V}{\pro \bH^2}\Big|_{\bH=\bar \mu^2/g}
=\dfrac{g^2}{\bar g^2}   \nn\\
=1+\dfrac{11g^2}{24 \pi^2} \Big(\ln \dfrac{\bar \mu^2}{\mu^2}
-c+\dfrac32 \Big),
\end{gather}
and retrieve the well known $\beta$-function which explains 
the asymptotic freedom \cite{wil}
\bea
\beta(\bar\mu)= \bar\mu \dfrac{\pd \bar g}{\pd \bar\mu}
= -\frac{11 \bar g^3}{24\pi^2}.
\label{betaf}
\eea
This confirms that our calculation is consistent with the known 
result.

In terms of the running coupling the renormalized potential 
is given by
\bea
V_{\rm ren}=\dfrac12 \bH^2
\Big[1+\dfrac{11 \bar g^2}{24 \pi^2 }
(\ln\dfrac{\bar g \bH}{\bar\mu^2}-\dfrac{3}{2})\Big],
\eea
which generates a non-trivial local minimum at 
\bea
\langle \bH \rangle=\dfrac{\bar \mu^2}{\bar g} 
\exp\Big(-\dfrac{24\pi^2}{11\bar g^2}+ 1\Big).
\eea
The corresponding effective potential is plotted in Fig. \ref{2pot}
where we have assumed $\bar \alpha_s = 1$ and $\bar \mu =1$. 
This is nothing but the monopole condensation which generates 
the desired mass gap and dimensional transmutation in QCD.  

To summarize, the gauge invariant and parity conserving 
monopole background and the color reflection invariance
allows us to demonstrate the stable monopole condensation 
and the generation of the mass gap in SU(2) QCD gauge 
independently. In particular, the color reflection invariance
(the C-projection) removes the tachyonic modes which 
destabilized the Savvidy vacuum. 

Notice that the monopole background could also contain 
the electric (i.e., Coulomb) part.  And we can generalise 
the above calculation with an arbitrary chromo-electromagnetic 
monopole background. To do so we choose an arbitrary 
$\bar H_\mn$ which has constant electric and magnetic 
fields $\bar E$ and $\bar H$ given by 
\bea
&\bH=\dfrac12 \sqrt{\sqrt{\bH_\mn^4
+(\bH_\mn \tilde{\bH}_\mn)^2}+\bH_\mn^2},  \nn\\
&\bE=\dfrac12 \sqrt{\sqrt{\bH_\mn^4
+(\bH_\mn \tilde{\bH}_\mn)^2}-\bH_\mn^2}.
\eea 
With this we can integrate out the chromon pair gauge 
invariantly, imposing the color reflection invariance. 
 
Since the color reflection invariance assures that the two spin 
polarizations of the chromon have exactly the same contribution 
in the effective action, we have \cite{prd02,jhep05,prd13}
\bea
&\Delta S = i \ln {\rm Det} [(-\bD^2+g\bH)(-\bD^2+g\bH)] \nn\\
&+i \ln {\rm Det} [(-\bD^2-ig \bE)(-\bD^2-ig \bE)] \nn\\
&- 2i \ln {\rm Det}(-\bD^2),   \nn\\
&\Delta {\cal L} =  \lim_{\epsilon\rightarrow0}
\dfrac{g^2}{8 \pi^2}  \Int_{0}^{\infty} \dfrac{dt}{t^{1-\epsilon}} 
\dfrac{\bH \bE}{\sinh (g\bH t/\mu^2) \sin (g\bE t/\mu^2)} \nn\\
& \times \Big[\exp(-2g\bH t/\mu^2)
+\exp(+2ig\bE t/\mu^2)-1 \Big].
\label{eaabo}
\eea
This is the correct integral expression of SU(2) QCD effective 
action. 

Integrating this we have
\bea
{\cal L}_{eff}=\left\{\begin{array}{ll}-\dfrac12 \bH^2
-\dfrac{11g^2 }{48\pi^2}\bH^2
(\ln \dfrac{g\bH}{\mu^2}-c'),~~\bE=0 \\
~\dfrac12 g^2 \bar E^2 +\dfrac{11g^2}{48\pi^2} \bE^2
(\ln \dfrac{g \bE}{\mu^2}-c') \\
-i\dfrac{11g^2}{96\pi} \bE^2,
~~~~~~~~~~~~~~~~~~~~~~~~~~~~\bH=0  \end{array}\right.
\label{ceaabo}
\eea
Notice that when $\bH=0$ it has a negative imaginary 
part, which implies the pair annihilation of chromons \cite{prd02,jhep05,sch}. This must be contrasted with 
the QED effective action where the electron loop generates 
a positive imaginary part \cite{schw,prl01}. This difference 
is a direct consequence of the Bose-Einstein statistics 
of the chromon loop. Of course the quark loop in QCD, due to 
the Fermi-Dirac statistics, will generate a positive imaginary part 
which diminishes the asymptotic freedom \cite{prd02,jhep05}.

This has a deep meaning. The positive imaginary part in 
QED means the pair creation which generates the screening. 
On the other hand in QCD we must have the anti-screening 
to explain the asymptotic freedom, and the negative 
imaginary part is exactly what we need for the asymptotic 
freedom \cite{prd02,jhep05,sch}.  

The effective action (\ref{ceaabo}) has an important 
symmetry, the electric-magnetic duality \cite{prd02}. 
We can obtain the two effective actions for $\bH=0$ 
and $\bE=0$ from each other simply with the following 
replacement
\bea
\bE \rightarrow i\bH,~~~~~~\bH \rightarrow -i\bE.
\label{dt}
\eea
This duality was first discovered in the QED effective 
action \cite{prl01}. But subsequently this duality has 
been shown to exist also in the QCD effective action \cite{prd02,jhep05}. 

This tells that the duality should be regarded as a fundamental 
symmetry of the effective action of gauge theory, Abelian and 
non-Abelian. The importance of this duality is that it provides 
a very useful tool to check the self-consistency of the effective 
action. The fact that the two effective actions are related by 
the duality assures that they are self-consistent. Notice
that this duality is different from the well known duality 
in Maxwell's theory, that the theory is invariant under 
$\bE \rightarrow \bH,~\bH \rightarrow -\bE$. 

\section{One-loop Effective Action of SU(3) QCD and Monopole 
Condensation}

To obtain the one-loop SU(3) effective action we follow 
the same procedure. Make the Abelian decomposition and 
integrate out the colored valence gluons $\X_\mu$ (three 
chromons $\W_\mu^p$) gauge invariantly with the monopole 
field as the classical background, imposing the color reflection invariance. What is really remarkable is that the Weyl 
symmetry of the Abelian decomposition greatly simplifies 
this procedure. 

From the Weyl symmetric Lagrangian ({\ref{3ecd}) we have
\begin{gather}
\exp \big[iS_{eff}(\hA_\mu)\big]
=\Int \Pi_p \cD \W_\mu^p \cD \vc^p \cD \vc^{p*} \nn\\
\exp \Big\{i\Int \Big[\sum_p \Big\{-\dfrac{1}{6} (\hF_\mn^p)^2
-\dfrac{1}{4} (\hD_\mu^p \W_\nu^p- \hD_\nu^p \W_\mu^p)^2 \nn\\
-\dfrac{g}{2} \hF_\mn^p \cdot (\W_\mu^p \times \W_\nu^p) \Big\} 
-\sum_{p,q} \dfrac{g^2}{4} (\W_\mu^p \times \W_\mu^q)^2 \nn\\
-\sum_{p,q,r} \dfrac{g}2 (\hD_\mu^p \W_\nu^p- \hD_\nu^p \W_\mu^p)
\cdot (\W_\mu^q \times \W_\mu^r)  \nn\\
-\sum_{p\ne q} \dfrac{g^2}{4} \big((\W_\mu^p \times \W_\nu^q)
\cdot (\W_\mu^q \times \W_\nu^p)  \nn\\
+(\W_\mu^p \times \W_\nu^p)\cdot (\W_\mu^q 
\times \W_\nu^q) \big) \nn\\
+\sum_p \vc^{p*} \bD_\mu D_\mu \vc^p
-\dfrac{1}{2\xi} \sum_p(\bD_\mu^p \W_\mu^p)^2 \Big] d^4 x \Big\},
\label{3ea0}
\end{gather}
where we have imposed the gauge fixing condition
$\hat D_\mu \vec X_\mu =\sum_p(\bD_\mu^p \W_\mu^p)=0$. 

At the first glance the integral looks complicated, but there are 
two things which simplify the integral. First, in the one loop 
approximation only the terms quadratic in $\W^p_\mu$ contribute 
to the integral. Second, the Weyl symmetric Abelian decomposition 
(\ref{3ecd}) reduces the chromon functional integral to the sum of 
three SU(2) integral of $\W_\mu^p$. 

So the integral expression of the effective action is simplified to
\begin{gather}
\exp \big[iS_{eff}(\hA_\mu)\big]
\simeq \sum_p \Int \cD \W_\mu^p \cD \vc^p \cD \vc^{p*} \nn\\
\exp \Big\{i\Int \Big[-\dfrac{1}{6} (\hF_\mn^p)^2
-\dfrac{1}{4} (\hD_\mu^p \W_\nu^p- \hD_\nu^p \W_\mu^p)^2 \nn\\
-\dfrac{g}{2} \hF_\mn^p \cdot (\W_\mu^p \times \W_\nu^p)  \nn\\
+ \vc^{p*} \bD_\mu D_\mu \vc^p
-\dfrac{1}{2\xi} (\bD_\mu^p \W_\mu^p)^2 \Big] d^4 x \Big\}.
\label{3ea0}
\end{gather}
This effectively reduces the calculation of the SU(3) QCD 
effective action to that of SU(2) QCD calculation. In general 
this applies to the SU(N) QCD effective action, because 
the Weyl symmetry holds in the Abelian decomposition of 
any SU(N) QCD. This is why the calculation of the SU(2) QCD effective action is so important. This simplification would 
have been impossible without the Weyl symmetric Abelian decomposition. 

Now, all we have to do is to add the SU(2) result (\ref{ceaabo}) 
in a Weyl symmetric way. With the constant monopole 
background $\bH_\mn^p$ given by
\begin{gather}
\bH_p = \dfrac12 \sqrt{\sqrt{(\bH_\mn^p)^4
+ (\bH_\mn^p \tilde \bH_\mn^p)^2} + (\bH_\mn^p)^2},  \nn\\
\bE_p = \dfrac12 \sqrt{\sqrt{(\bH_\mn^p)^4
+ (\bH_\mn^p \tilde \bH_\mn^p)^2} - (\bH_\mn^p)^2}, 
\end{gather}
we have
\bea
&\Delta S = 2i \sum_p \ln {\rm Det} 
\big[(-\bD_p^2+2g\bH_p)(-\bD_p^2-2ig \bE_p) \big] \nn\\
&- 2i \sum_p \ln {\rm Det}(-\bD_p^2),
\label{fdabx3}
\eea
so that
\begin{gather}
\Delta {\cal L} =  \lim_{\epsilon\rightarrow0}
\sum_p  \dfrac{g^2}{8 \pi^2} \Int_{0}^{\infty} 
\dfrac{dt}{t^{1-\epsilon}} \dfrac{\bH_p \bE_p}{\sinh (g\bH_p t/\mu^2)
\sin (g\bE_p t/\mu^2)}  \nn\\
\times \Big[ \exp(-2g\bH_p t/\mu^2) 
+\exp(+2ig \bE_p t/\mu^2)-1 \Big].
\label{eaab3}
\end{gather}
Notice that for the chromo-magnetic background we have 
$\bE_p=0$, but for the chromo-electric background we have 
$\bH_p=0$.

From this we have the following explicitly Weyl symmetric 
effective Lagrangian. For $\bE_p=0$ we have
\begin{gather}
{\cal L}_{eff} =- \sum_p \Big(\dfrac{\bH_p^2}{3}
+\dfrac{11 g^2}{48\pi^2} \bH_p^2
(\ln \dfrac{g\bH_p}{\mu^2}-c) \Big), 
\label{3eaa}
\end{gather}
and for $\bH_p=0$ we have
\begin{gather}
{\cal L}_{eff} =\sum_p \Big(\dfrac{\bE_p^2}{3}
+\dfrac{11 g^2}{48\pi^2} \bE_p^2
(\ln \dfrac{g\bE_p}{\mu^2}-c) \nn\\
- i \dfrac{11 g^2}{96\pi} \bE_p^2\Big).
\label{3eab}
\end{gather}
Just as in SU(2), the effective action has a negative imaginary 
part when $\bH_p=0$. This again tells that the chromo-electric 
field annihilates the chromon pairs, which implies the anti-screening 
and asymptotic freedom \cite{prd02,jhep05,prd13,wil}. 
Moreover, the effective action has the dual symmetry. 
It is invariant under the dual transformation 
$\bH_p \rightarrow -i\bE_p$ and $\bE_p \rightarrow i\bH_p$. 

We can express the effective Lagrangians (\ref{3eaa}) and 
(\ref{3eab}) in terms of three Casimir invariants $C_p$ of 
SU(3). For example, for the pure chromo-magnetic background 
$\bH_\mn^a=\bH_\mn n^a+\bH'_\mn {n'}^a$ we have 
\begin{gather}
C_1=(\bH_\mn^a)^2, \nn \\
C_2=(d^{abc} \bH_\mn^b \bH_\mn^c)^2, \nn \\
C_3=(d^{abc} \bH_\mn^a \bH_{\nu\rho}^b 
\bH_{\rho \mu}^c)^2,
\end{gather}
which are related to $\bH_p$ by the following identities
\begin{gather}
\bH_1^2+\bH_2^2+\bH_3^2=\dfrac32 C_1, \nn \\
\bH_1 \bH_2 + \bH_2 \bH_3+\bH_3 \bH_1 
=\dfrac34 C_1^2-\dfrac{9}{16} C_2, \nn \\
\bH_1 \bH_2 \bH_3 =\dfrac18C_1^3-\dfrac{3}{16} C_1 C_2 
-\dfrac{3}{2} C_3.
\label{eqs}
\end{gather}
So, expressing $\bH_p$ by $C_p$ we can replace $\bH_p$ 
in (\ref{3eaa}) by $C_p$. Similarly, for the pure chromo-electric 
(Coulombic) background we have exactly the same identities,
with $\bH_p$ replaced by $\bE_p$. This assures that we can 
express the effective Lagrangian by three Casimir invariants.

\begin{figure}
\psfig{figure=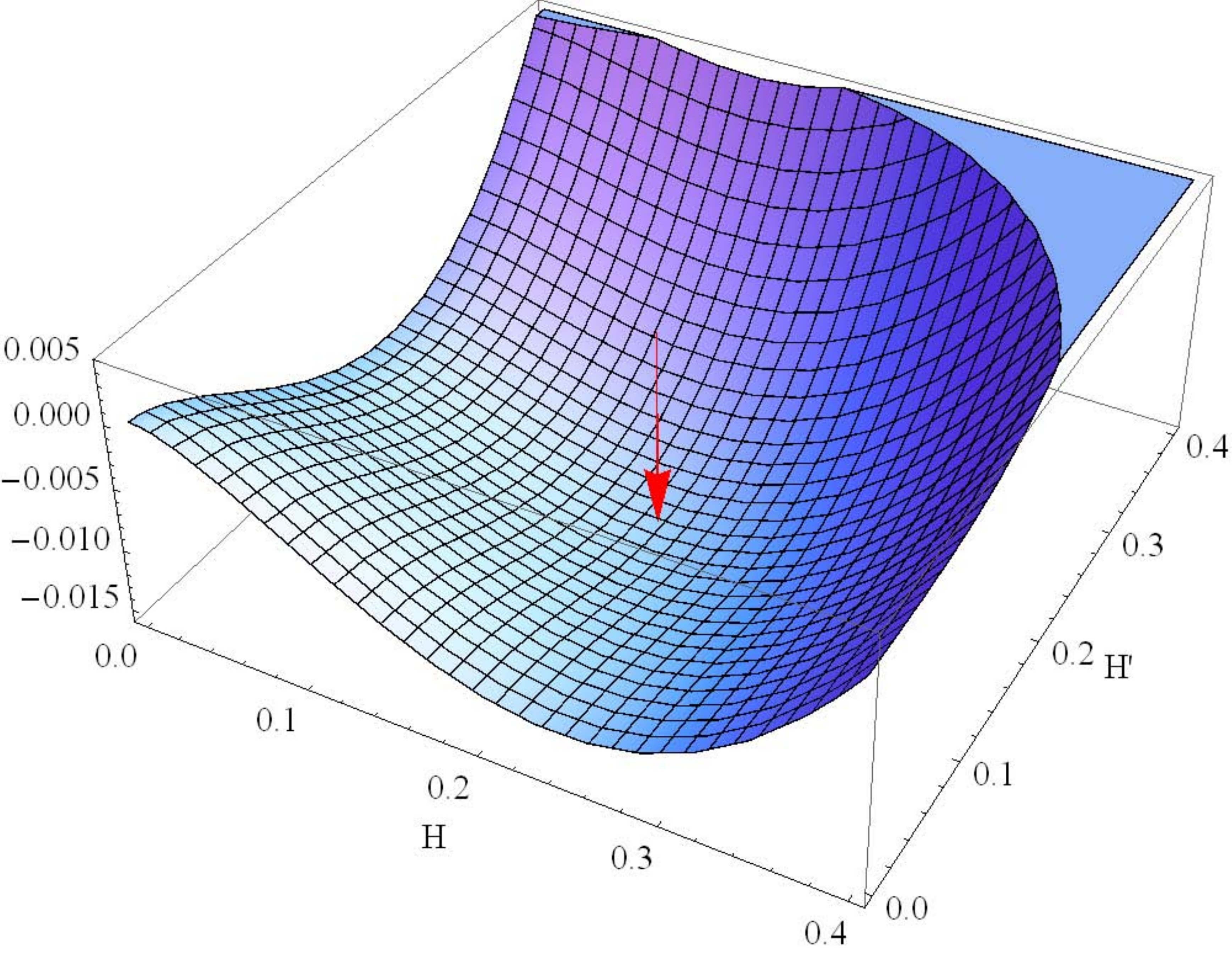, height = 4.5cm, width = 6.5 cm}
\caption{\label{3pot1} The QCD effective potential
with $\cos \theta =0$, which has a unique minimum at 
$H=H'=H_0$.}
\end{figure}

Notice that the classical potential depends on only one Casimir 
invariant $C_1=\bH_1^2+\bH_2^2+\bH_3^2$, but the effective 
potential depends on all three Carsimir invariants (or equivalently 
three independent variables $\bH_1,~\bH_2$, and $\bH_3$). 
On the other hand, the chromo-magnetic background in SU(3) 
QCD is given by seemingly two independent monopole 
fields $\vH$ and $\vH'$. So we need to understand what is 
the origin of the third degree. 

To understand this notice that $\vH$ and $\vH'$ in principle 
can have different space orientation, so that the angle $\theta$
which describes the relative orientation of two vectors $\vH$ 
and $\vH'$ in real space can be arbitrary. So the constant 
chromo-magnetic background has three degrees, $\bH$, 
$\bH'$ and $\theta$. In fact from (\ref{cp3}) we have
\begin{gather}
\bH_1=|\vH|,~~~\bH_2=\bH_+,~~~\bH_3=\bH_-,   \nn\\
H_\pm=|\dfrac12 \vH\pm \dfrac{\sqrt 3}{2} \vH|  \nn\\
=\sqrt{\dfrac14 \bH^2+\dfrac34  \bH'^2
\pm \dfrac{\sqrt 3}{2} \bH \bH' \cos \theta},  \nn\\
\cos \theta=(\vH \cdot \vH')/\bH \bH'.
\end{gather} 
This shows that the classical background indeed has three 
degrees of freedom. 

With the magnetic background (with $\bE=0$) we have 
the effective potential given by
\begin{gather}
V_{eff}=\sum_p \Big(\dfrac{\bH_p^2}{3}
+\dfrac{11 g^2}{48\pi^2} \bH_p^2
(\ln \dfrac{g\bH_p}{\mu^2}-c) \Big).
\label{effpot}
\end{gather}
We can renormalize the potential by defining a running 
coupling $\bar g^2(\bar \mu^2)$
\begin{gather}
^{\forall_p}~\dfrac{\pro^2 V_{eff}}{\pro \bH_p^2}
\Big |_{\bH_1=\bH_2=\bH_3=\bar \mu^2}  
=\dfrac{g^2}{\bar g^2}   \nn\\
= 1+ \dfrac{11 g^2}{16 \pi^2} (\ln \dfrac{\bar \mu^2}{\mu^2}
-c+\dfrac{5}{4}),
\label{renorm}
\end{gather}
from which we retrieve the SU(3) QCD $\beta$-function
which assures the asymptotic freedom \cite{wil}
\bea
& \beta (\bar \mu) = \bar \mu \dfrac{d \bar g}{d \bar \mu}
=-\dfrac{11 \bar g^3}{16 \pi^2}.
\label{beta}
\eea
With this we have the renormalized potential 
\begin{gather}
V_{eff}=\sum_p \Big(\dfrac{\bH_p^2}{3}
+\dfrac{11 \bar g^2}{48\pi^2} \bH_p^2
(\ln \dfrac{g\bH_p}{\bar \mu^2}-c) \Big).
\label{renpot}
\end{gather}
We plot the effective potential for $\cos \theta=0$ in 
Fig. \ref{3pot1} and for $\cos \theta =1$ in Fig. \ref{3pot2} 
for comparison, where we have put $\bar \mu=1$ and 
$\bar \alpha_s=1$.

The potential has the absolute minimum at 
$\bH_1=\bH_2=\bH_3=H_0$ (or equivalently 
$\bH=\bH'=H_0$ and $\cos \theta =0$), 
\begin{gather}
\langle \bH_1\rangle=\langle \bH_2\rangle
=\langle \bH_3\rangle =\dfrac{\bar \mu^2}{\bar g}
\exp \big(-\dfrac{16\pi^2}{11 \bar g^2} 
+\dfrac34 \big) =H_0,   \nn\\  
V_{min} = -\dfrac{11 \bar \mu^4}{32 \pi^2} 
\exp \big(-\dfrac{32\pi^2}{11 \bar g^2}+\dfrac{3}{2} \big).
\end{gather}
This is the monopole condensation, or more precisely
the monopole-antimonopole pair condensation,  which 
generates the dimensional transmutation and the mass 
gap in SU(3) QCD.

\begin{figure}
\begin{center}
\psfig{figure=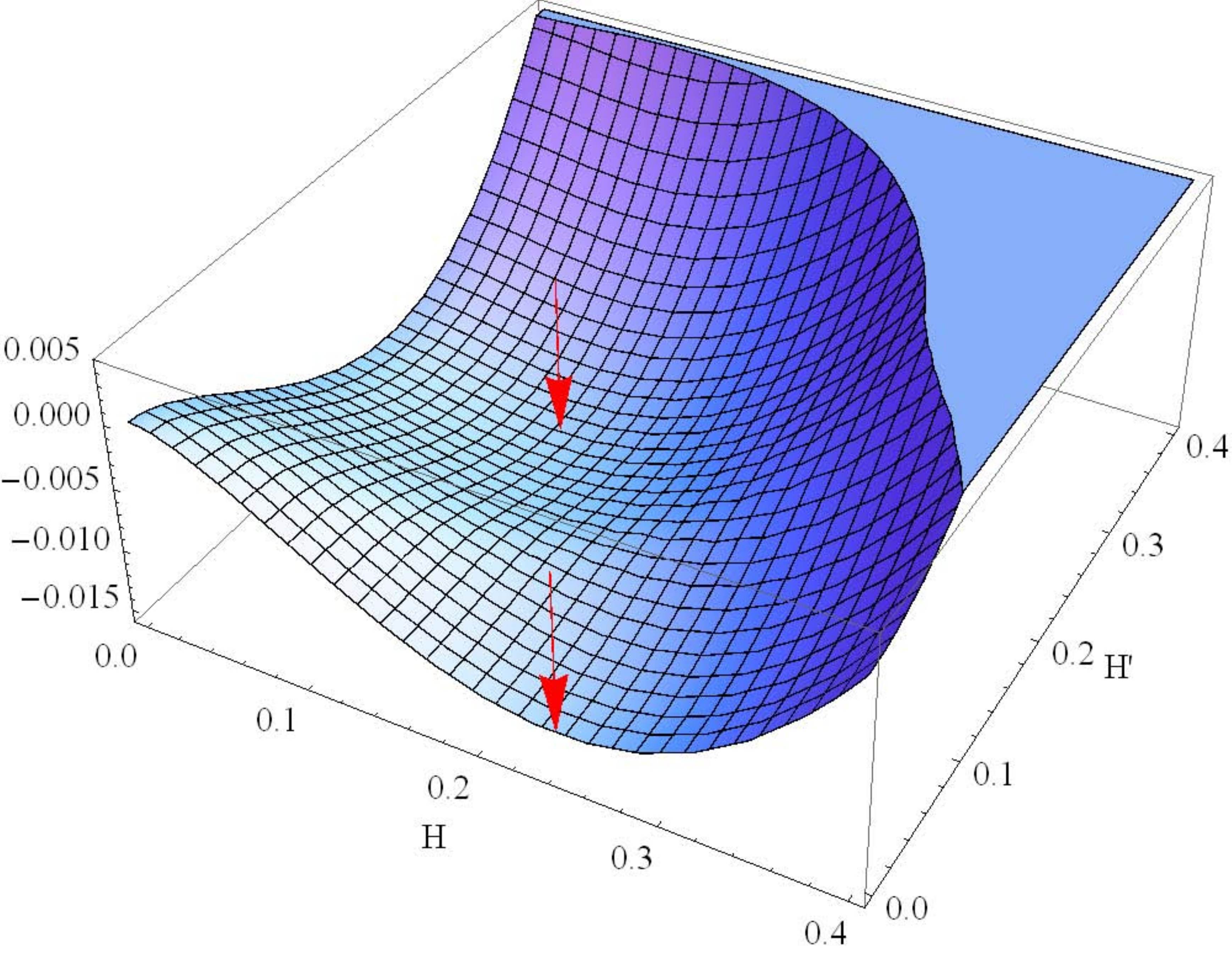, height = 4.5cm, width = 6.5 cm}
\end{center}
\caption{\label{3pot2} The effective potential with 
$\cos \theta =1$, which has two degenerate minima.}
\end{figure}

Notice that when $\vec H$ and $\vec H'$ are parallel 
(i.e., when $\cos \theta=1$) it has two degenerate
minima at $\bH=2^{1/3} H_0,~\bH'=0$ and at
$\bH=2^{-2/3} H_0,~\bH'= \sqrt3 \times 2^{-2/3} H_0$.
This is shown in Fig. \ref{3pot2}. In terms of 
$(\bH_1,\bH_2,\bH_3)$ the two degenerate minima are 
given by $(2^{1/3}H_0, 4^{-1/3}H_0, 4^{-1/3}H_0)$
and $(4^{-1/3}H_0, 2^{1/3}H_0, 4^{-1/3}H_0)$.

It must be emphasized that the Weyl invariance plays a crucial 
role in the true SU(3) QCD vacuum. To see this, notice that 
the effective potential (\ref{renpot}) is Weyl symmetric. So it 
is natural to assume that the minimum point has the maximal 
Weyl symmetry. This implies that the minimum point must 
form a singlet under the Weyl transformation, or $\bH_1
=\bH_2=\bH_3$. This is fulfilled when $\vH$ and $\vH'$ 
are orthogonal and $\bH=\bH'$. And this is exactly the true 
minimum shown in Fig. \ref{3pot1}.  

The above analysis demonstrates the followings. First, 
of course, SU(3) QCD has the stable monopole condensation 
which could be identified as the physical vacuum. Second, 
the monopole condensation naturally reproduces the asymptotic 
freedom. Third, the chromo-electric flux makes the pair 
annihilation of chromons. This confirms that essentially 
all qualitative features of the SU(2) QCD effective action 
translate to the SU(3) QCD effective action, or in general
SU(N) QCD effective action. 

Obviously, just as in SU(2) QCD, the color reflection 
invariance plays the crucial role in SU(3) QCD. It is this 
symmetry which assures the the gauge invariance of 
the effective action and the stability of the monopole 
condensation. In retrospect this is exactly what we expected. 
Clearly the gauge invariance forbids colored objects from 
the physical spectrum in QCD, which is why we have the color 
confinement. But this gauge invariance, after the Abelian 
decomposition, becomes the color reflection invariance. 
So it is just natural that the color reflection invariance 
generates the monopole condensation which explains 
the color confinement. 

On the other hand, there are new features in the SU(3) effective 
Lagrangian. First, the effective Lagrangian is characterized by 
three variables $\bH_p$. This is understandable because SU(3) 
has three Casimir invariants. What is unexpected is that 
the monopole condensation given by $\vH$ and $\vH'$ (or
$\vE$ and $\vE'$) in general can have different space orientation, 
and only when $\vH$ and $\vH'$ are orthogonal and we have 
the true vacuum. As we pointed out, this is because the vacuum 
must have the maximal Weyl symmetry.

\section{Discussion}

In this paper we have shown how to generalize the calculation 
of the SU(2) QCD effective action to that of the SU(3) QCD, 
with the help of the Weyl symmetric Abelian decomposition. 
Our result confirms that all essential features of the SU(2) 
QCD effective action remain unchanged in the SU(3) QCD. 
In particular we have the stable monopole condensation which 
generates the dimensional transmutation and the mass gap. 

Our calculation should be contrasted with the old calculations 
which had critical defects \cite{savv,niel,ditt,yil}. The Savvidy 
vacuum has an intrinsic instability. Worse, it is not gauge 
invariant nor parity conserving. So it could not be the QCD 
vacuum. The Abelian decomposition allows us to remove 
these defects and obtain the stable monopole condensation.     

In specific the Abelian decomposition gives us the following 
advantages. First, it decomposes QCD to the classical 
background and the quantum fluctuation gauge independently. 
In particular, it allows us to separate the gauge invariant 
and parity conserving monopole background gauge 
independently. Second, it reduces the non-Abelian gauge 
symmetry to the simple and discrete color reflection 
symmetry. This makes the implementation of the gauge 
invariance much easier. Third, it allows us to express 
the SU(3) QCD, or in general SU(N) QCD, in terms of 
the SU(2) QCD in a Weyl symmetric way. This effectively 
reduces the calculation of the SU(3) QCD or in general 
SU(N) QCD effective action to that of the SU(2) QCD effective 
action. These are the new features which were lacking in 
the old calculations.
  
But we emphasize that the monopole condensation should 
really be understood as the monopole-antimonopole 
condensation. This is because in QCD the monopole and 
anti-monopole are gauge equivalent, since they are related 
by the gauge transformation \cite{prd80,prl81,plb82}.

This has a deep meaning. It has often been claimed that 
the color confinement in QCD comes from ``the dual Meissner 
effect" generated by the monopole condensation \cite{nambu}. 
We emphasize, however, that the confinement mechanism 
in QCD is not exactly dual to the Meissner effect which 
confines the magnetic flux in ordinary superconductor. 

In superconductor the magnetic flux is screened by 
the supercurrent generated by the electron pairs, without 
the positron pairs. And obviously the Cooper pairs carry 
the electric charge. But in QCD the chromo-electric flux 
is confined by the monopole-antimonopole condensation,
not by the supercurrent of monopole pairs. Besides,
the monopole-antimonopole pairs have no magnetic 
charge. Moreover, the underlying mass generation 
mechanism in the Meissner effect is the spontaneous 
symmetry breaking (i.e., Higgs) mechanism. But in QCD 
it is the dynamical symmetry breaking mechanism without 
any ad hoc input mass scale. This tells that the two 
confining mechanisms are not exactly dual to each other. 

An important consequence of this monopole-antimonopole 
condensation is that in QCD the colored flux which confines 
the $q\bar q$ pairs has no sense of helicity. As a result 
hadrons in the quark model have no parity doubling partners. 
This resolves the long standing problem of the parity 
doubling in hadron spectroscopy \cite{nc74}.  

Moreover our result confirms that in the presence of 
chromo-electric background the effective action (\ref{3eab}) 
has a negative imaginary part. This tells that the cheromo-electric 
flux has an instability which induces the pair annihilation, not 
the pair creation, of the chromons \cite{prd02,jhep05,wil,sch}. 
This should be compared with the pair creation of electrons 
in the electric background in QED, which makes the screening 
of the electric charge \cite{prl01,schw}. The negative imaginary 
part in (\ref{3eab}) tells that in QCD we have the anti-screening 
of color, which explains the asymptotic freedom.  

Obviously the proof of the monopole condensation in QCD 
is very important from the theoretical point of view. But one 
may ask if there is any way to verify this monopole condensation 
experimentally. There might. To see this consider the Meissner 
effect in superconductor characterized by two scales, 
the correlation length of Cooper pair and the penetration 
length of magnetic field. Physically they describe the masses 
of the Higgs field and massive vector field. In other words, 
the spontaneous symmetry breaking generates two physical 
states which can be verified experimentally.     

In QCD we could think of similar scales and similar phenomenon.
For example, we could think of the correlation length of 
the monopole-antimonopole pair and the penetration length 
of the color flux, which could create two new states, 
the ``magnetic" glueballs. In this logic we could have 
the $0^{++}$ and $1^{++}$ modes of the vacuum fluctuation 
of the monopole condensation, whose masses are fixed by 
two scales \cite{prd80,prl81}. 

On the other hand, the situation is different in QCD. Here we 
have a dynamical symmetry breaking which is characterized 
by one scale, $\Lambda_{QCD}$. Nevertheless, it is quite 
possible that the monopole condensation may have the vacuum 
fluctuation. This strongly suggests that, although the monopole 
condensation may not generate two physical modes, it could 
generate one vacuum fluctuation mode which can be identified 
as $0^{++}$ monoball \cite{prd13,prd15,prd18}. The experimental 
verification of such vacuum fluctuation mode could be viewed 
as a direct evidence of the monopole condensation.  

The Abelian decomposition does many things. In the perturbative 
regime it decomposes the gluons to neurons and chromons, 
and decomposes the Feynman diagram. In non-perturbative 
QCD it proves the monopole dominance and demonstrates 
the monopole condensation. But it has another deep impact 
on hadron spectroscopy. It generalizes the quark model to 
the quark and chromon model which provides a new picture 
of hadrons, in particular the glueballs. 

The identification of glueballs has been a big issue in QCD. 
The general wisdom is that QCD must have the glueballs 
made of gluons \cite{frit,jaff,coyne}. But the search for 
the glueballs has not been so successful for two reasons. 
First, theoretically there has been no consensus on how 
to construct the glueballs from the gluons. This made it 
difficult for us to predict what kind of glueballs we can 
expect. 

The other reason is that experimentally it is not clear how 
to identify the glueballs. This is partly because the glueballs 
could mix with the quarkoniums, so that we must take 
care of the possible mixing to identify the glueballs
experimentally \cite{prd15}. This is why we have very
few candidates of the glueballs so far, compared to huge
hadron spectrum made of quarks \cite{pdg}.

The Abelian decomposition provides a clear picture of glueball.
First, it separates the colored chromons from the color neutral 
neurons gauge independently. Second, it tells that the chromons, 
just like the quarks, become the colored constituent of hadrons 
while the neurons play the role of the binding gluons. So we can 
identify the glueballs as the color reflection invariant bound 
states of chromons. This naturally generalizes the quark model 
to the quark and chromon model, and provides a simple picture 
of the quarkonium-glueball mixing. This helps us to resolve 
the long standing glueball problem and identify the glueballs 
experimentally \cite{prd80,prl81,prd15,prd18}.    

This should be compared with two popular models of glueballs, 
the bag model and the constituent gluon model. The bag model 
identifies the glueball as gluon field confined in a bag \cite{jaff}. 
On the other hand the constituent gluon model identifies 
the glueballs as gauge invariant combinations of SU(3) color 
octet gluons \cite{coyne}. On the surface our quark and chromon 
model looks similar to the constituent gluon model, but is 
fundamentally different in one important respect. They treat 
all gluons on the same footing, with no distinction between 
the binding gluons and the constituent gluons. But our 
model treats only the chromons as the constituent gluons \cite{prd15,prd18}  

Before we close, we emphasize that the Abelian decomposition 
is not just a theoretical proposition. It can be tested by experiment. 
There are various ways, but a straightforward and direct way is 
to confirm the existence of two types of gluon jets, the neuron 
jet and chromon jet. As we have seen, the neuons behave like 
photons in QED, while the chromons behave like the quarks. 
This tells that the neuron jet should look like the photon jet, 
but the chromon jet should look like the quark jet. 

As we have pointed out, we already have enough knowledge 
on how to differentiate the gluon jet from the quark jet 
experimentally \cite{jet1,jet2}. Moreover, there has been 
a new proposal on how to separate different types of jets 
at LHC \cite{jet3}. Using these knowledges we could 
actually confirm the existence of two types of gluon jets 
experimentally. If confirmed, this could be more important 
than the confirmation of the gluon jet which established 
the asymptotic freedom \cite{wil}.          

In this paper we have neglected the quarks. We simply remark 
that the quark loop, just like the electron loop in QED, tend to 
diminish the asymptotic freedom. This is because the quark 
loop, unlike the chromon loop, generates the quark pair creation 
which makes the color screening. The reason is simple. The quark 
loop obeys the Fermi statistics but the chromon loop obeys 
the Bose statistics, so that the imaginary part they create in 
the chromo-electric background should have opposite signature.    

But if the number of quarks are small enough, the asymptotic 
freedom holds. In fact we can show that exactly the same 
constraint on the number of quarks is needed to keep 
the asymptotic freedom. 

Finally one could try to calculate the two-loop effective action 
of QCD. Certainly it goes without saying that the Abelian 
decomposition could also simplify the calculation greatly. 
On the other hand, the two-loop correction is not expected 
to change the one-loop result in any qualitative way. The details 
of the QCD effective action which includes the quark loop and 
the two-loop correction will be published elsewhere \cite{cho}.

{\bf Acknowledgements}

~~~The work is supported in part by National Research 
Foundation of Korea funded by the Ministry of Education 
(Grants 2015-R1D1A1A0-1057578 and 2018-R1D1A1B0-7045163), 
and by the Center for Quantum Spacetime at Sogang University.


\begin{thebibliography}{99}
\bibitem{r} Our notation of $\lambda$-matrices is the same as 
the GellMann's notation, except that our $\lambda_5$ has 
the opposite signature. Our notation is more convenient for 
the Weyl symmetry.
\end{thebibliography}
\end{document}